\documentclass[twocolumn]{article}
\usepackage[
top=2.0cm,
bottom=2.0cm,
left=1.8cm,
right=1.8cm,
columnsep=0.6cm
]{geometry}
\usepackage{amsmath}
\usepackage{amssymb}
\usepackage{booktabs}
\usepackage{lineno}

\usepackage{fancyhdr}
\usepackage{graphicx,natbib}
\usepackage{authblk}
\usepackage{sectsty}
\usepackage[normalem]{ulem}
\usepackage[
colorlinks=true,
linkcolor=blue,
citecolor=blue,
urlcolor=blue,
breaklinks=true
]{hyperref}
\usepackage{xcolor}
\usepackage{subcaption}
\usepackage{orcidlink}

\newcommand{\abs}[1]{\left|#1\right|}

\begin{document}
\title{Generative AI for image reconstruction in Intensity Interferometry: a first attempt}

\author{Km Nitu Rai\thanks{E-mail: niturai201296@gmail.com}\orcidlink{0000-0002-9939-805X}}
\affil[1]{\small{\em{Aryabhatta Research Institute of Observational Sciences, Manora Peak, Nainital 263129, India.}}}
\affil[2]{\small{\em{School of Physics, Indian Institute of Science Education and Research Thiruvananthapuram, Maruthamala PO, Vithura, Thiruvananthapuram 695551, Kerala, India.}}}
	
\author[3]{Yuri van der Burg}
\affil[3]{\small{\em{Physik-Institut, University of Zurich,	Winterthurerstrasse 190, 8057 Zurich, Switzerland.}}}	

\author[2]{Soumen Basak\orcidlink{0000-0003-4042-0387}}

\author[3]{Prasenjit Saha\orcidlink{0000-0003-0136-2153}}

\author[4, 5]{Subrata Sarangi\orcidlink{0000-0001-6500-9987}}
\affil[4]{\small{\em{School of Applied Sciences, Centurion University of Technology and Management, Odisha-752050, India.}}}
\affil[5]{\small{\em{Visiting Associate, Inter-University Centre for Astronomy and Astrophysics, Post Bag 4, Ganeshkhind, Pune 411 007, Maharashtra, India.}}}

\setbox0=\vbox{\hsize=6.5truein
	\hbox to \hsize{\hss\bf Abstract\hss}
	\smallskip \noindent 
	
	In the last few years Intensity Interferometry (II) has made significant strides in achieving high-precision resolution of stellar objects at optical wavelengths. Despite these advancements, phase retrieval remains a major challenge due to the nature of photon correlation. This paper explores the application of a conditional Generative Adversarial Network (cGAN) to tackle the problem of image reconstruction in II. This method successfully reconstructs the shape, size, and brightness distribution of simulated, fast-rotating stars based on sparsely sampled spatial power spectra obtained by using two different hypothetical ground-based II facilities composed of six and nine Imaging Atmospheric Cherenkov Telescopes (IACTs), respectively.  Although this particular example could also be addressed using parameter fitting, our results suggest that with larger arrays of IACTs much more complex systems with varied surface features could be reconstructed by applying machine-learning techniques to II. Hence this approach merits closer examination.}

\date{\box0}

\maketitle


\section{Introduction}
Humans instinctively feel a symbiotic relationship with the stars. One of the primary scientific projects of humanity is to figure out what the stars are and how do they do what they do. The first obvious step in this project, beyond measuring their global parameters like diameter, mass, orbital and astrometric elements, is to obtain images of the stars with all details of the stellar surfaces. In case of the Sun, this is done routinely by observatories and Sun-observation satellites. But such a routine still remains a challenge even for the $\alpha$-Centauri system, our nearest stellar neighbour. The objective is to achieve capability of high fidelity image reconstruction of distant stars. Two interferometry based techniques, namely, Michelson Interferometry (MI) and Intensity Interferometry (II) have emerged during the last century to address this objective. A discussion of the development of these two approaches and comparison of their respective merits and challenges can be found in, for example, \cite{Rai2025}. The work reported here presents the results of a first effort at applying a Conditional Generative Adversarial (neural) Network (cGAN) to image reconstruction of a fast rotator using its simulated II observations.  

The foundational basis of II stems from the pioneering experiments and theoretical investigations carried out initially by Hanbury Brown and Twiss (HBT) \citep{HBT56Lab,brown1957interferometry, brown1958interferometry} and, later, by \cite{glauber1963quantum} and others. The correlation between photons emitted by a thermal source within a certain bandwidth (known as ``photon bunching'', also widely referred to as the \textquotedblleft HBT effect\textquotedblright) measured by a pair of photon detectors in two partially coherent beams of light was reported in 1956 by \cite{HBT56Lab}. \cite{Rai2025} present a recent variant of this experiment, carried out with pseudo-thermal light. The HBT Effect and the related theoretical investigations laid the foundation for the modern field of Quantum Optics.

Hanbury Brown and his collaborators led the creation and installation of the historic II facility at Narrabri, Australia, and reported the measurement of angular diameters of 32 stars and a few multiple star-systems \citep{hanbury1974angular}. Soon after this work, however, II observations of stars was stalled for over four decades due to the limits of the then available photon detectors and data processing equipment. With gradual mitigation of such issues, proposals to utilize Imaging Atmospheric Cherenkov Telescope (IACT) facilities for conducting II observations of stars have emerged\citep{LeBohec2006, nunez2010stellar, nunez2012high, 2013APh....43..331D} as a secondary science application of these facilities during moonlit nights. SII observations at VERITAS, MAGIC, and H.E.S.S. are now being reported \citep[e.g.,][]{2024ApJ...966...28A,2024MNRAS.529.4387A,2025MNRAS.537.2334V}. This approach has the potential to enhance the scientific output of existing IACT facilities and the upcoming ones such as the Cherenkov Telescope Array Observatory (CTAO). Simulations \citep[e.g.,][]{10.1093/mnras/stab2391, 10.1093/mnras/stac2433} have shown that recent advancements in photon detectors hold promising prospects of high-precision stellar parameter measurements by such facilities.

The thrust of such efforts has been to measure average and global parameters  such as angular diameters, binary separations, and orbital characteristics of stars and star systems. High-fidelity imagery, however, would transcend such integrated view of the systems and offer direct insights into dynamic stellar surface phenomena, including limb darkening, convection cells, granulation, star spots, oblateness and gravity darkening in rapid rotators and atmospheric structures, akin to the detailed observations routinely conducted on our own Sun.

As it stands today, studies grappling with various issues of image reconstruction are being reported \citep{Haubois2009, Norris2021AZCyg, Liu2024SuperresolutionII, Liu2025}. Projects of MI-based image reconstruction have made substantial progress as demonstrated in the cases of Betelgeuse \citep{Haubois2009} and AZ~Cyg \citep{Norris2021AZCyg}. On the other hand, II-based methods are at a nascent stage. Recently Liu et al. have demonstrated \citep{Liu2024SuperresolutionII, Liu2025} through outdoor experiments using ``active optical intensity interferometry'', imaging of millimeter-scale targets located at km scale distances and achieving resolution ~14 times better than a single telescope’s diffraction limit. A \textquotedblleft flexible computational algorithm\textquotedblright reconstructs images from intensity correlations, overcoming atmospheric turbulence and optical imperfections.

We report here, the results of our attempt -- the first of its kind -- to construct the gravity-darkened sky-images of fast rotating stars consistent with their respective simulated ground-based II-observations using a cGAN neural network architecture. Image reconstruction of gravity-darkened fast-rotating stars has long been examined using various methods in MI \citep{vanBelle2001, DomicianodeSouza2003, DomicianodeSouza2005, mcalister2005first, Monnier2007, Pedretti2009, Zhao2009, Martinez2021}. Recently photosphere oblateness of $\gamma$-Cassiopeia \citep{Archer-arXiv-2025} has been measured at the VERITAS observatory using II. These results put our work in context, and our work presented here is a natural next step especially of the work by \cite{Archer-arXiv-2025}. We implement a cGAN model \citep{isola2017image} to reconstruct images of fast-rotating stars using their simulated Intensity Interferograms and sky-intensity distributions as input data for training, testing, and validation. The intensity interferograms are synthetically generated by simulations of the fast rotators using two hypothetical IACT facilities having, respectively, arrays of $N_T\, =\, 6$ and $N_T\, =\, 9$ Cherenkov telescopes. The images predicted by the trained cGAN shows promising results in reconstructing the stars' shapes and sizes. The reconstructed brightness distributions are then assessed using moments.

This paper is organized as follows. The next section discusses briefly the past efforts at image reconstruction using II, followed by a section that discusses the signal and noise characteristics in II observations of fast rotating stars (FRSs) using the IACT facilites with $N_T$ number of telescopes. An outline of the method to incorporate the effect of Earth's rotation during the observation is also presented. The following section introduces the cGAN formulation and its structure. The fifth section details the parameter selection for training the network for the task of image reconstruction. The sixth section presents the results of the trained network both visually and via image moments. Finally, the paper concludes with a discussion of the overall results.

\section{Past Efforts at Image Reconstruction relevant to Intensity Interferometry}
Several approaches have been developed for phase reconstruction in intensity interferometry (II). \cite{gamo1963triple} introduced triple-intensity correlation, applied by \cite{goldberger1963use} to microscopic systems and extended by \cite{sato1978imaging, sato1979computer, sato1981adaptive} to measure stellar diameters and phases. The results of this work were limited by low signal-to-noise ratio (SNR). \cite{GerchbergSaxton1972} proposed an iterative phase retrieval method using image and diffraction plane data. This approach was sensitive to initial estimates and had  to struggle with convergence speed issues. \cite{Fienup1982} improved this approach with a Hybrid Input-Output algorithm, enhancing robustness in noisy conditions. \cite{holmes2010two} utilized Cauchy-Riemann relations for 1-D and 2-D image reconstruction, applied to simulated stellar data with Imaging Cherenkov Telescope Arrays \citep{nunez2010stellar, nunez2012high, nunez2012imaging}, but faced computational complexity for higher dimensions. \cite{Li2014} introduced a regularized iterative method incorporating priors (e.g., sparsity) to mitigate noise and ill-posedness, though challenged by parameter tuning and initial guess sensitivity. The Transport-of-Intensity Equation (TIE), proposed by \cite{Teague1983}, retrieved phase from intensity variations across the planes. \cite{Zhang2020} solved TIE as a Poisson equation using a maximum intensity assumption, while \cite{Kirisits2024} combined TIE with the Transport of Phase Equation for improved accuracy across arbitrary apertures and non-uniform illumination, with convergence dependent on initial guesses and boundary conditions.

With non-linearity built into their architecture, artificial neural networks (ANNs) empowered by deep learning methods  present a promising machine-learning based method for exploring the task of reconstructing images of stellar objects from ground-based observations. Convolutional Neural Networks (CNNs), with their specialized architecture for processing two-dimensional datasets, are a natural choice for image processing tasks. In astronomical image reconstruction projects, a common challenge is that the data from the interferometric plane are typically sparsely sampled and are noisy as well. Therefore, the CNN architectures and the deep learning methods employed must be capable of reliably learning both the global context of the training dataset and the local features within it. Among the various CNN architectures, U-Net models \citep{ronneberger2015u} have proven successful in such tasks.

Furthermore, given that achieving a high signal-to-noise ratio (SNR) is often challenging in astronomical datasets, it is immensely beneficial if additional data can be generated using the available information from the observed sky density distribution and ground-based observations (II data, in our case) of the sources under investigation. Generative Adversarial Networks (GANs), introduced by \cite{goodfellow2014generative}, have been successful in such data augmentation tasks. Conditional GAN (cGAN) architectures, proposed by \cite{mirza2014conditional} and applied to a wide variety of datasets by \cite{isola2017image}, leverage additional information about the images in the training datasets and have demonstrated remarkable robustness in image recovery across diverse data types.

In the astrophysical context, \cite{schawinski2017galaxypics} employed a GAN model to recover features, such as spiral arms, central bulges, and disk structures of galaxies, from noise-affected images. \cite{mustafa2019cosmogan} developed and customized a Deep Convolutional GAN, dubbed \textquotedblleft CosmoGAN\textquotedblright, capable of generating high-fidelity weak-lensing convergence maps of dark matter distribution that statistically reproduce real weak lensing structures. \cite{coccomini2021lightweightgan} have successfully generated credible images of planets, nebulae, and galaxies using \textquotedblleft lightweight\textquotedblright\ and \textquotedblleft physics-uninformed\textquotedblright\ GANs to produce synthetic images of celestial bodies. They also generated a \textquotedblleft Hubble Deep Field-inspired\textquotedblright\ wide-view simulation of the universe. 

Encouraged by the success of machine-learning approaches, especially GAN-based structures, across various areas of Astrophysics, we have attempted to adapt this technique to the task of image reconstruction in II.

\section{Intensity Interferometry (II) with IACT arrays}

\begin{figure*}
	\centering
	\includegraphics[width=0.49\linewidth]{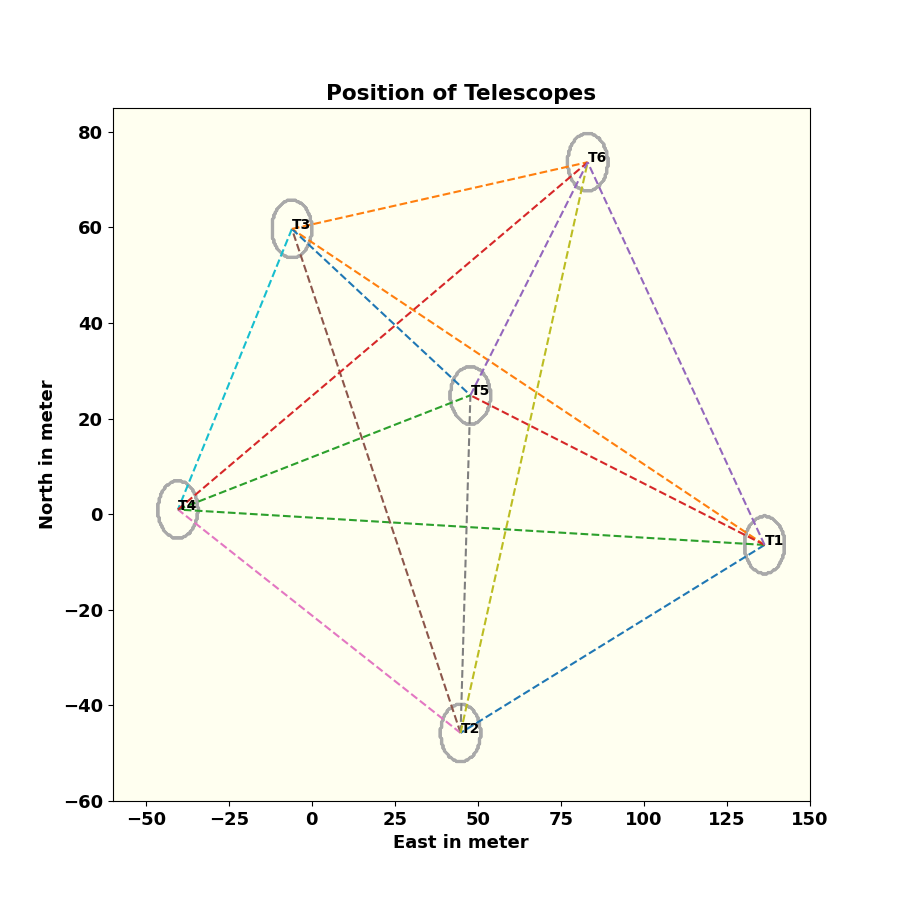}\hfil
	\includegraphics[width=0.49\linewidth]{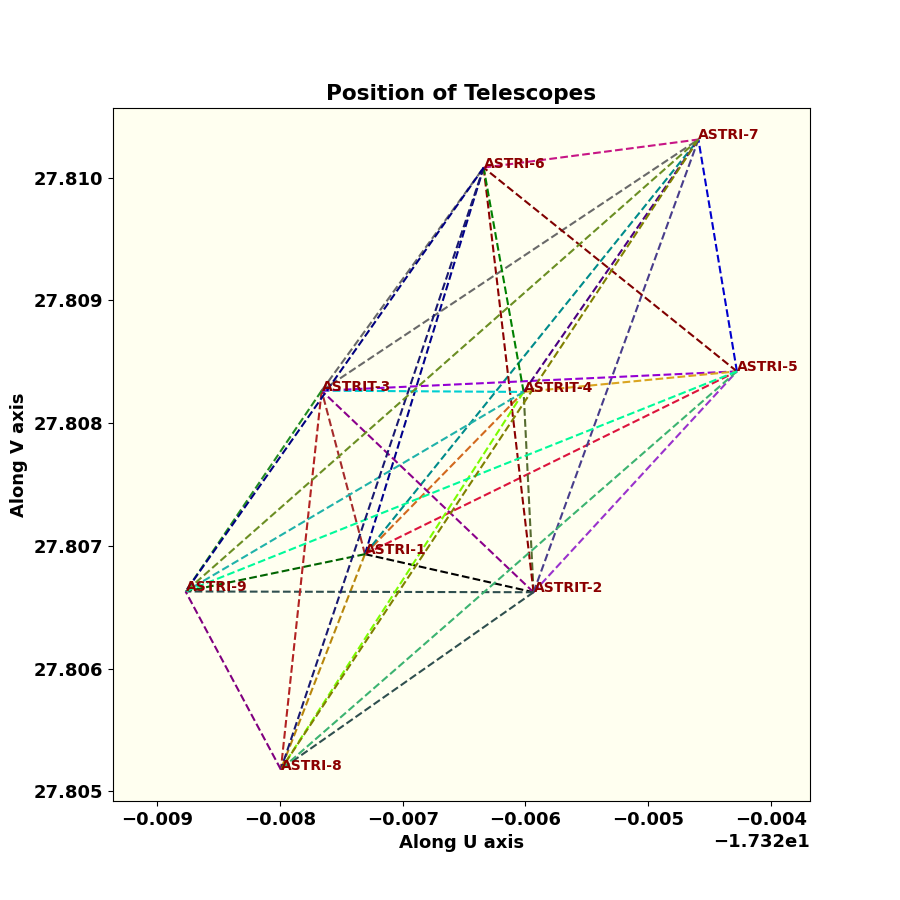}
	\caption{A schematic representation of a hypothetical observation facility with an array of $N_T = 6$  Imaging Atmospheric Cherenkov Telescopes (IACTs) (left panel) and $N_T = 9$ IACTs that mimic the telescope configuration of ASTRI \cite{scuderi2022astri} (right panel). In the work presented here, these two sets of arrays are used in simulating the II observation of the fast rotators.}
	\label{fig:teles}
\end{figure*}

\begin{figure*}
  \includegraphics[width=0.49\linewidth]{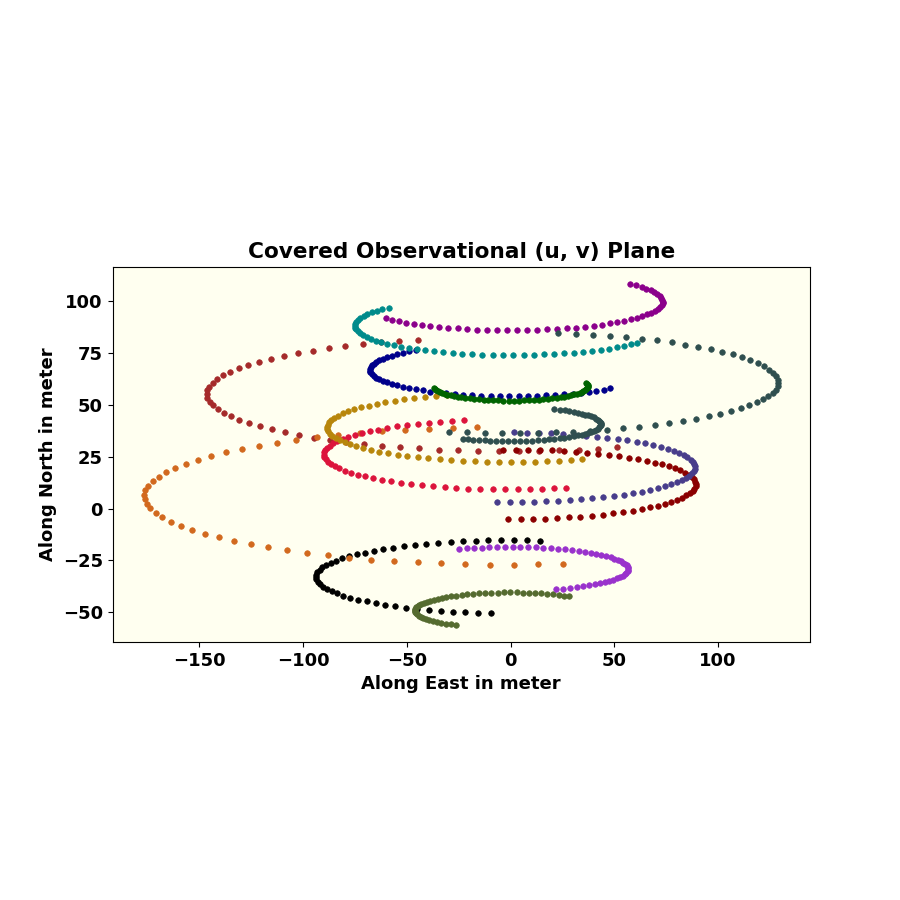}\hfil
  \includegraphics[width=0.49\linewidth]{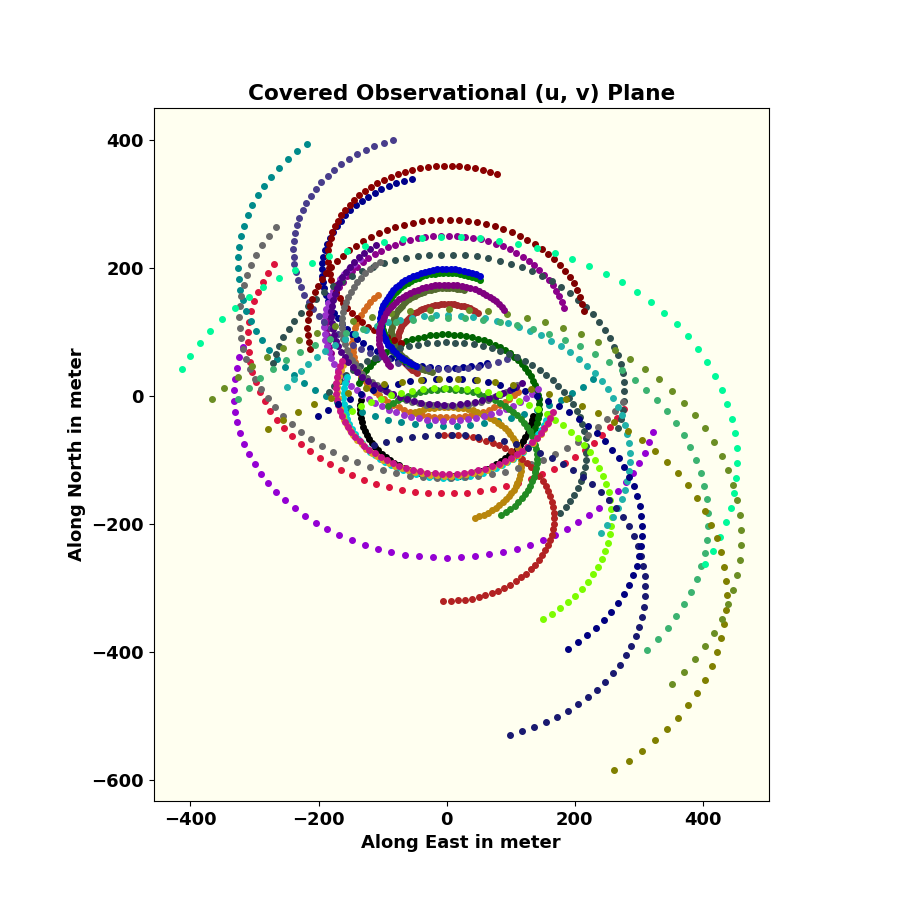}
  \caption{The tracks of the baselines of the two hypothetical II observation facilities shown in fig.~\ref{fig:teles} generated (synthetically) due to Earth's rotation during one night of simulated II observation. The number of baselines scales as square of number of telescopes in the array thus leading to greater coverage of the observational plane and better image reconstruction prospects.} 
\label{fig:base}
\end{figure*}

\begin{figure}[hbt]
  \includegraphics[width=\linewidth]{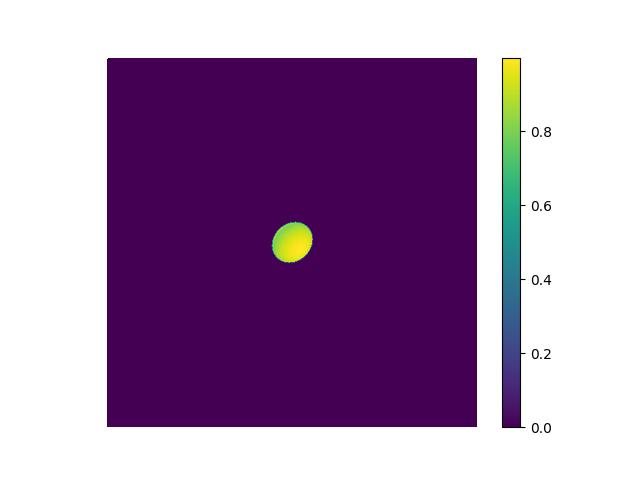}
  \caption{This figure shows one of the simulated fast rotating stars (FRS). Brightness is the highest at the poles; gravity darkening is visible along the equator. A total of 31460 such images of FRS with different parameter values have been generated to train the model.}
  \label{fig:image}
\end{figure}
\begin{figure*}
	\includegraphics[width=.33\linewidth]{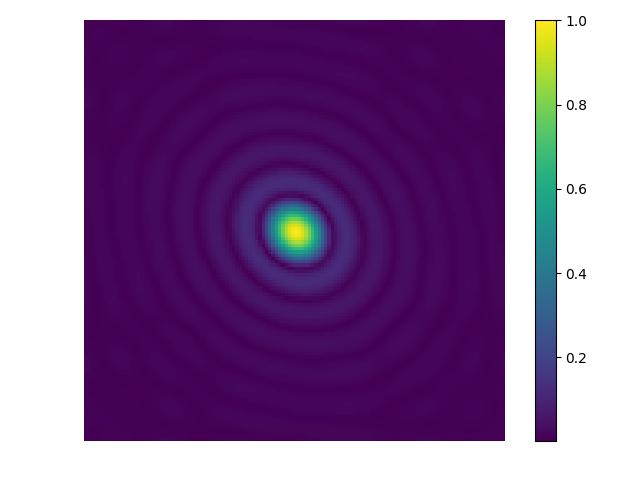}\hfil
	\includegraphics[width=.33\linewidth]{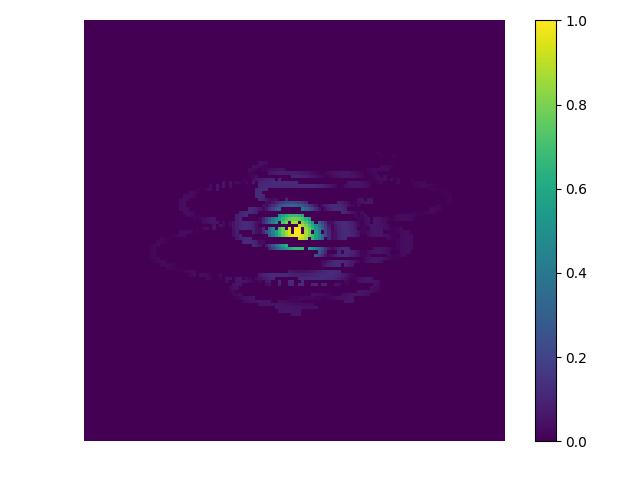}\hfil
	\includegraphics[width=.33\linewidth]{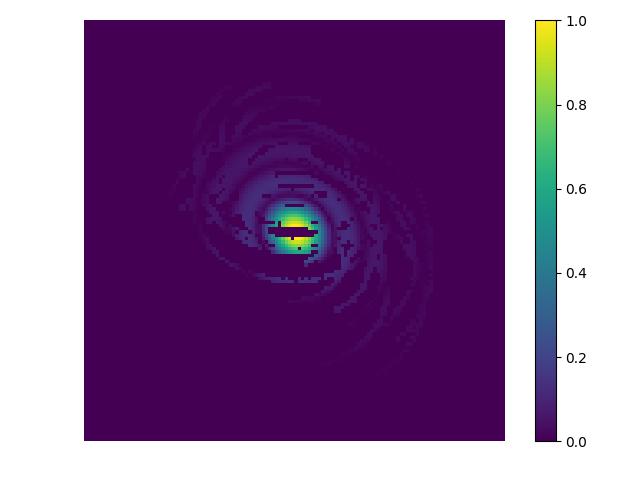}
	\caption{The left panel shows the absolute value of the two-dimensional Fast Fourier Transform (FFT) of the source depicted in Fig.~\ref{fig:image}. It represents the interferometric ($u,v$) plane intensity distribution that would be generated by an infinite number of baselines (or an infinite number of telescopes) observing the source. The middle and right panels represent the same intensity distribution as sampled along the tracks shown in the left and right panels of Fig.~\ref{fig:base} respectively. The middle and left panels reflect the sparse nature of the signal received by a realistic finite number of telescopes and baselines sampled from the full ($u,v$) plane signal space. All figures are plotted on a linear scale and normalized to the maximum pixel value in each respective figure.}
	\label{fig:ft}
\end{figure*}

This section presents a brief conceptual overview of how an array of telescopes is used to perform II observations, and explains the SNR from these measurements.
\subsection{The signal for II}\label{sec:signal}
As a simple example, let us consider an identical pair of IACTs pointed at a star. Suppose the two telescopes simultaneously measure the intensity of radiation $I_1(t)$ and $I_2(t)$, respectively. The signals from these detectors are cross-correlated and averaged over time, yielding the second order ($n=2$) correlation of these intensities as \citep[cf.][]{acciari2020optical, 2013APh....43..331D}
\begin{equation}
	g^{(2)}_{12} = \frac{\left\langle I_1(t) \cdot I_2(t + \tau) \right\rangle}{\langle I_1(t) \rangle \cdot \langle I_2(t) \rangle} 
	\label{eqn:HBT}
\end{equation}
where $\tau$ is the time delay between the telescopes. For spatially coherent and randomly polarized light, Eq.~\eqref{eqn:HBT} reduces to the relation \citep[sometimes called the Siegert relation, see e.g.,][]{acciari2020optical}.
\begin{equation}
	g^{(2)}_{12} = 1 + \frac{\Delta f}{\Delta \nu} \abs{V_{12}}^2
	\label{eqn:HBT2}
\end{equation}
where $\Delta f$ is the electronic bandwidth of the photon detectors which measure the intensities and $\Delta \nu$ is the frequency bandwidth of the filters employed in the telescopes to observe the star.  Values of $\Delta\nu\sim 1\,\mathrm{THz}$ and $\Delta f \sim 1\,\mathrm{GHz}$ are typical of recent work.  In Eq.~\eqref{eqn:HBT2}, $V_{12}$, referred to as the complex visibility function, is the Fourier transform of the source brightness distribution. It contains information about the star's angular diameter. However, the phase information is lost since we measure only the absolute value $\vert V_{12} \vert^2$. In observational astronomy, the correlation is often expressed in terms of the normalized contrast $c_{12}$, given by:
\begin{equation}
    c_{12} = g^{(2)}_{12} - 1 = \frac{\Delta f}{\Delta \nu} \abs{V_{12}}^2
	\label{eq:signal}
\end{equation}
with $\abs{V_{12}}^2$ being a function of baseline $b = \sqrt{u^2 + v^2}$ on the observational plane $(u, v)$. This implies the strength of the signal would be enhanced if a larger number of baselines or pairs of telescopes are employed. 

The total correlation obtained by all the pairs of $N_T$ telescopes in an observation facility, excluding the zero-baseline cases, becomes
\begin{equation}
c_{N_T} = \sum_{\substack{i,j =1 \\ i<j}}^{N_T}  c_{ij}.
\label{eqn:total-corr}
\end{equation}

\subsection{The Signal-to-Noise Ratio for II}
The primary purpose of IACT arrays is to observe the Cherenkov showers generated in the upper atmosphere by high-energy gamma rays (with energy $E\ \geq 30$ GeV) arriving from cosmic sources. When employed for II observations, these telescopes focus light received from a sky source onto their respective set of photo-multiplier tubes (PMTs), see e.g., \citep{aleksic2016major} with appropriate specifications. In the simulation models adopted here, we consider, in succession, two arrays of IACTS each with similar properties. The positional configurations of these arrays of $N_T=6$ and $N_T =9$ IACTs are respectively shown in the left and the right panels of Fig.~\ref{fig:teles}. The optical signal directed to a PMT is filtered using a spectral filter with a chosen mean observational wavelength $\lambda$ and corresponding bandpass $\Delta \lambda$. The use of filters not only reduces background noise but also improves the signal quality and the efficiency of the PMTs. Filtering background skylight becomes even more significant in II observations, as, currently, these are carried out during bright moon-lit nights when the primary function of the IACTs (of observing Cherenkov Showers) is rendered infeasible. It is important to note that the light from the stellar source is focused on a PMT attached with each of the telescopes during II observations.

The significance of the signal can be expressed in terms of the signal-to-noise ratio (SNR), which depends on many factors. However, most importantly, it does not depend on the optical bandwidth $\Delta {\mathrm {\nu}}$ of the radiation for a two-telescope correlation. The explanation for the independence of the SNR from $\Delta {\mathrm {\nu}}$ is provided in several works \citep[e.g., subsection 4.1 of][]{10.1093/mnras/stab2391}. The Signal-to-Noise Ratio ($\mathrm{SNR_{12}}$) produced by the considered identical pair of IACTs becomes
\begin{equation}
	{\mathrm {SNR_{12}}} = A \cdot \alpha \cdot q \cdot n \cdot F^{-1} \cdot \sigma \cdot \sqrt{\frac{T \Delta f}{2}} \cdot \abs{V_{12}}^{2}
	\label{eq:SNR}
\end{equation}
Here, $A$ is the total mirror area, $\alpha$ is the quantum efficiency of the PMTs, $q$ is the throughput of the remaining optics, and $n$ is the differential photon flux from the source. The excess noise factor of the PMTs is represented by $F$, $T$ denotes the observation time, and $\sigma$ is the normalized spectral distribution of the light (including filters) \citep[see e.g.,][]{acciari2020optical}.

The total SNR obtained by all the pairs of $N_T$ telescopes in an observation facility excluding the zero-baseline cases, becomes
\begin{equation}
\mathrm{SNR}_{N_T} = \sum_{\substack{i,j =1 \\ i<j}}^{N_T}  {\mathrm {SNR}_{ij}}.
\label{eqn:total-SNR}
\end{equation}
Point to note in eqns~\ref{eq:SNR} and \ref{eqn:total-SNR} is that all the $N_T$ IACTs in the facility along with their associated optics and other parameters are considered to be identical.

\subsection{Baseline considerations}\label{sec:earth}
The measurement of the size of stellar objects via squared visibility depends on the distance between the telescopes, known as the baseline $b$.
\begin{equation}
	|V_{12}(b)|^2 = \frac{c(b)}{c(0)}
	\label{eq:angular_size_meas}
\end{equation}
For achieving a good SNR with a given telescope configuration, covering as much as possible of the interferometric plane is always desirable. If the source is at the zenith, the coordinates in the Fourier plane ($u,v$) are given by:
\begin{equation}
	(u,v) = \frac{1}{\lambda} (b_E, b_N)
\end{equation}
where $b_E$ and $b_N$ are, respectively, the baselines expressed in east and north coordinates. However, the sources can be anywhere on the sky, and the telescopes are stationary and may also have different relative altitudes $b_A$ depending on the available terrain. In order to gather maximum possible information on the source during the observation session and to cover as much of the observational plane as possible during such sessions, Earth's rotation must be taken into account using rotated baselines.  For a given stellar source with declination $\delta$ and hour-angle $h$, as observed by telescopes located at latitude $l$, equation (\ref{eq:baseline_rot}) provides the rotated baselines for a given pair of telescopes \citep[see e.g., eqs.~8--10 from][]{2020MNRAS.498.4577B} with the $R$-matrices representing the respective rotation operations.
\begin{equation}
\begin{pmatrix} u\\ v\\ w\\ \end{pmatrix} = R_x(\delta) \cdot R_y(h) \cdot R_x(-l) \begin{pmatrix} b_E \\ b_N \\ b_A \\ \end{pmatrix}
	\label{eq:baseline_rot}
\end{equation}

Fig.~\ref{fig:base} shows, respectively, the tracks of 15 and 36 baselines generated, due to Earth's rotation from the telescopes shown in the left and the right panel of Fig.~\ref{fig:teles}. Since every pair of telescopes traces an ellipse in the Fourier plane, the total number of ellipses scales as
\begin{equation}
	\label{eq:N_telescopes}
	\mathcal{N} = \tfrac12 N_T \cdot (N_T -1)
\end{equation}
where $N_T$ is the number of telescopes considered.
As the number of baselines increases quadratically with $N_T$, Intensity Interferometry (II) benefits greatly from a large number of telescopes. The upcoming CTAO Global facility with upto 65 IACTs across both the hemispheres can offer many more baselines and play a major role in both $\gamma$-ray observations and II. \cite{2013APh....43..331D} has considered the telescope configurations being planned and has showed how it would provide a dense coverage of the interference plane.

\subsection{A Fictitious Fast Rotating Star: Our Test Case}
In our work presented here, we attempt image reconstruction of fast-rotating stars using their simulated Intensity Interferometric observation in a cGAN architecture. Fast-rotating stars are important test cases for understanding various astrophysical processes, including stellar evolution, internal structure, and dynamical behaviour over time. Fast rotation causes stars to adopt an oblate shape, flattening at the poles and bulging at the equator due to the stronger centrifugal force \citep[e.g.,][]{von1924radiative, 1999A&A...347..185M}. Fig.~\ref{fig:image} shows an image qualitatively representing a fictitious fast-rotating star, with brightness (and, therefore, the effective surface temperature) distributed across its surface. The brightness (effective temperature) is highest at the poles and lowest at the equator, a phenomenon known as gravity darkening \citep{lucy1967gravity}. First direct interferometric detection of stellar photospheric oblateness (of Altair) was pioneered by \cite{vanBelle2001} using the Palomar Testbed Interferometer (PTI) and the Navy Prototype Optical Interferometer (NPOI). Gravity darkening due to fast rotation was first observed through interferometric and spectroscopic data from CHARA Array for the fast-rotating star Regulus \citep{mcalister2005first}.  As pointed out earlier, these two pieces  of work, all using Michelson Interferometry, make a subset of several others \citep{vanBelle2001, DomicianodeSouza2003, mcalister2005first, DomicianodeSouza2005, Monnier2007, Pedretti2009, Zhao2009, Martinez2021}. The first observation of photospheric oblateness (of $\gamma$ Cassiopeiae or $\gamma$-Cas) using Intensity Interfereometry (II) has been recently carried out at VERITAS observatory and is reported by \cite{Archer-arXiv-2025}. Reportage of such observations of other $\gamma$-Cas like targets and other class of FRS by Cherenkov Telescope arrays, such as  the MAGIC array, are expected by 2026. In addition, observation and measurement of gravity darkening using II is the natural next step and is yet to be reported. In this context, our work of reconstructing the image of FRSs from their II-simulated observations using cGAN is an attempt at solving this inverse problem along with mitigation of loss of phase information in II. 

II counts the photons arriving at the telescopes from the stellar object. The correlation of photon arrivals at the telescopes yields the squared visibility Eq.~\eqref{eq:angular_size_meas}, as explained in subsection~\ref{sec:signal}. The left panel of fig.~\ref{fig:ft} shows the normalized ground intensity distribution signal from the source shown in Fig.~\ref{fig:image}. A point to note here is that this figure represents the signal from the source that would be recorded by an infinite number of baselines provided by an infinite number of telescopes on the interferometric plane. In practice, only a small part of this information is available (as seen in the middle and the right panels of Fig.~\ref{fig:ft}), because one has a finite number of baselines corresponding to the finite number $N_T$ of telescopes at our disposal and a limited observation schedule. We have simulated the II observation of the fictitious star by two hypothetical observation facilities having, respectively, arrays of $N_T=6$ and of $N_T=9$ IACTs with their relative positions (correlated with baselines as seen in left and right panel of Fig.~\ref{fig:teles}). The observation is carried over one night. Using this modest amount of signal from one night's observation, we have trained a cGAN to construct the images of the sources.
\begin{figure*}
   \centering
   \includegraphics[width=0.49\linewidth]{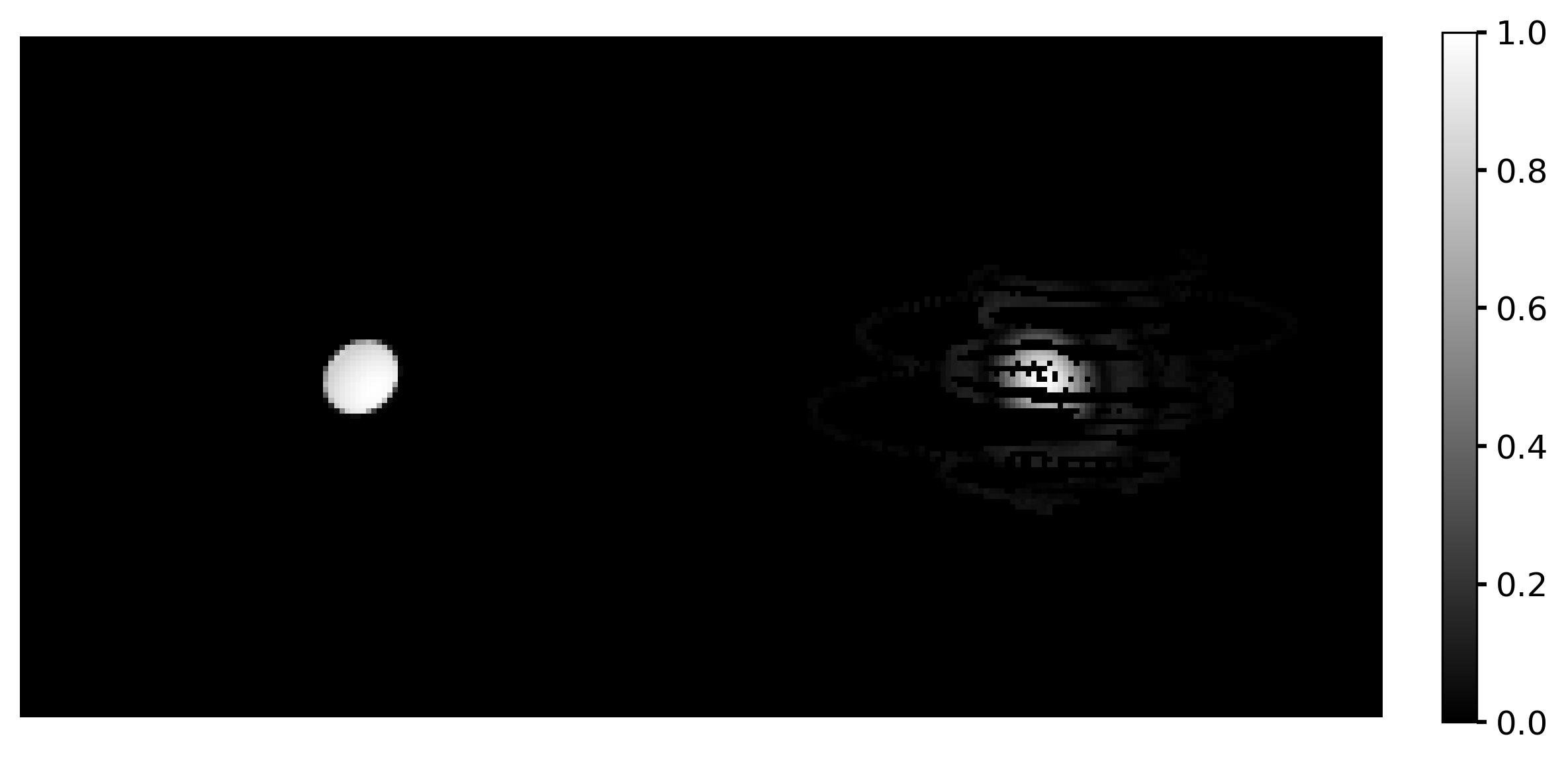}\hfil
   \includegraphics[width=0.49\linewidth]{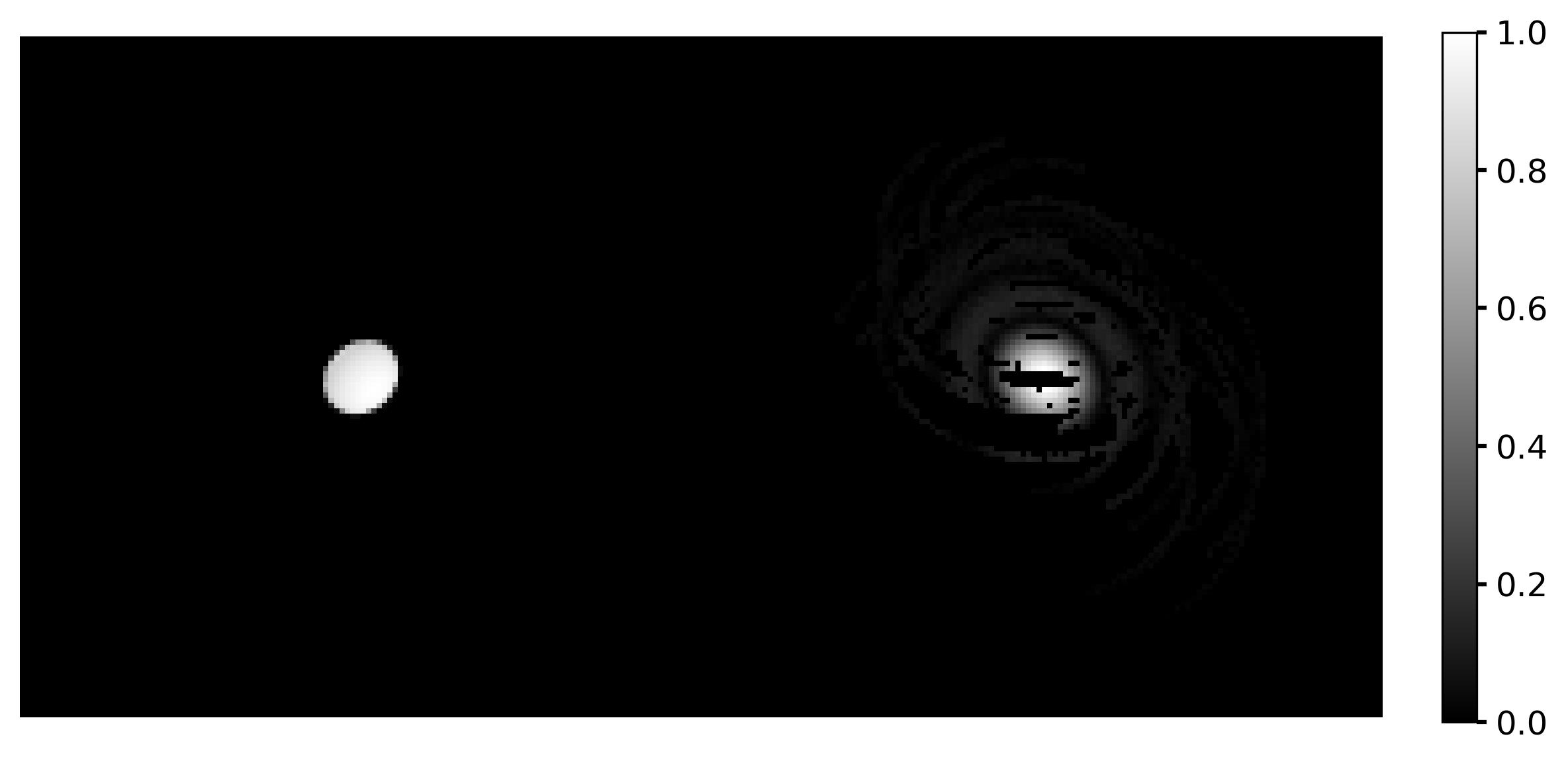}
   \caption{Two illustrative pieces of example of the input used for training the cGAN model. The picture on the left in both of the panels represents the source image in grey scale, which serves as the ground truth or the real data ($x$), as mentioned in the Flow Diagram (Fig.\ref{fig:FlowDiagram} discussed later). The picture on  the right, in both of the panels represents the simulated II observation pattern of the source image $x$. These two II observation patterns of the two panels are generated using the arrays of IACTs and the Earth's rotation. The pattern on the left (right) panel is generated using the array of $N_T = 6$ ($N_T =9$) IACTs. This pattern referred to as $y$, in the Flow Diagram (Fig.\ref{fig:FlowDiagram} discussed later) forms the ``condition" during the training to which the GAN model has to conform. Salt-and-pepper noise is added to this pattern for enhancing the robustness of the cGAN model. Together, the source image $x$ on the left and the II pattern $y$ on the right form a training pair. The GAN learns to construct the ``predicted'' image similar to ground truth $x$ from the noisy baseline signal $y$. The grayscale in both images is normalized to the brightest pixel.}
  \label{fig:GANinput}
\end{figure*}

\section{Generative Adversarial Networks}
\begin{figure*}
\centering
\includegraphics[width=\linewidth]{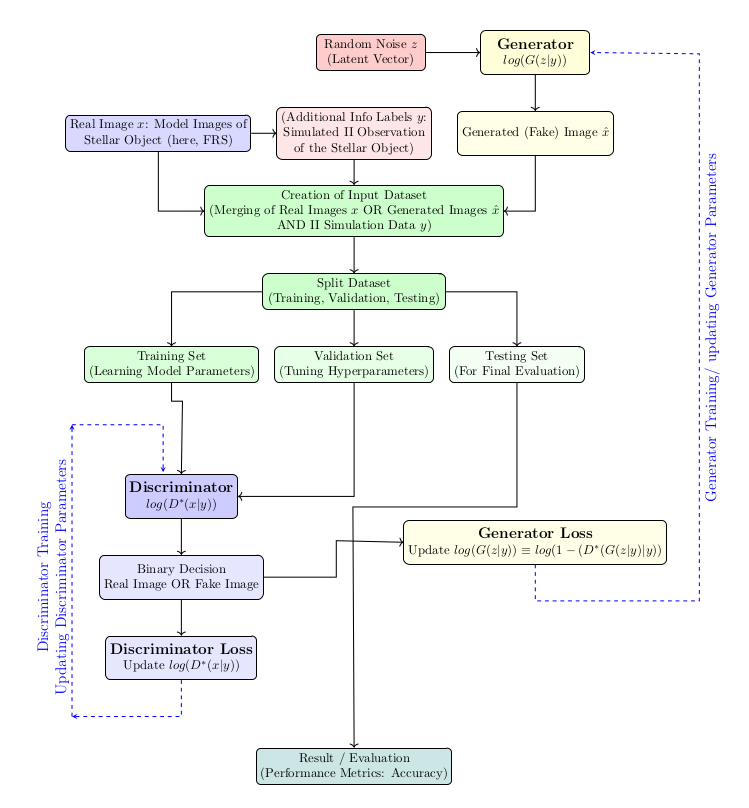}
\caption{A schematic representation of the features of the cGAN model used in this work and the iterative process of its training, validation and testing. The process constitutes four broad stages: (1) choice of an appropriate GAN architecture including both the Discriminator and the Generator (not shown in this figure) (2) preparation of the Training, Validation and Testing datasets and (3) Training and Validation of the Model (4) Testing and Evaluation of the Model. The stages (2), (3) and (4) are depicted in this figure. The datasets are prepared in three broad steps: (i) generating the ``ground truth'' images $x$ of fictitious fast rotating stars (FRS) , the sparse II observation signal $y$ used as the ``condition'' images in the Model and the generated images $\hat x$ sampled from a Normal Distribution (ii) merging these images into individual files with $(x|y)$ and/ or $(\hat x|y)$ (as seen in the two illustrated samples in Fig.(\ref{fig:GANinput})) and generating the full dataset in this process, and finally (iii) partitioning the full dataset into Training Set, Validation Set and the Testing Set. After the iterative training of the Model and its validation process is complete (``Nash point'' of the Minimax Game is reached), the Model is tested using the Testing Set and evaluated.}
\label{fig:FlowDiagram}
\end{figure*}
 
\begin{figure}
	\centering
	\includegraphics[width=0.9\linewidth]{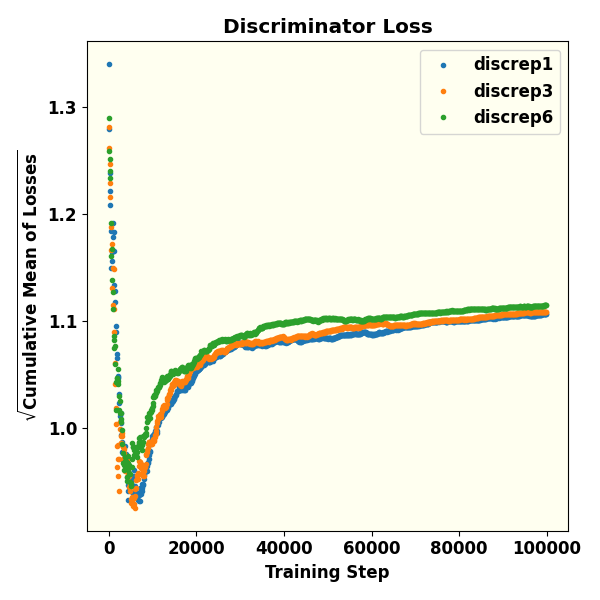}
	\includegraphics[width=0.9\linewidth]{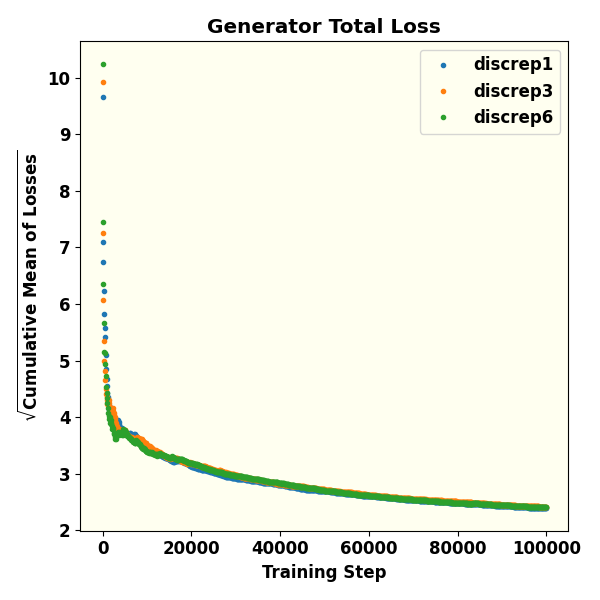}
	\caption{These figures show the effect of the ratio of episodes of Discriminator training per every episode of Generator training. This hyperparameter is termed Discriminator repetition or (discrep) in the figures. The square root of the cumulative mean of the losses are plotted against the training steps for 3 values of this ratio. Understandably, this ratio has a higher impact on the Discriminator loss than the Generator loss. Equal number of episodes of training produces minimum cumulative loss of the Discriminator. The dip in the Discriminator loss during the initial phases of its training can be interpreted as its early success in detecting ``fake" (or generated) images because of a poorly trained Generator. With gradual training of the Generator, the success rate of the Discriminator decreases and eventually approaches saturation with equal probability of being successful in telling ``fake" from real. The continual decrease and eventual saturation of the Generator loss is a result of its training to generate better images with increasing number of steps.}
	\label{fig:Plot_discrep_loss}
\end{figure}

Generative Adversarial Networks (GANs) were introduced by \cite{goodfellow2014generative}. A GAN model involves two competing deep neural network models, referred to as the Generator and the Discriminator. These two networks engage in a zero-sum ``Minimax” game, as in Game Theory. Given a real data set $\{x_i\}$ (for example, a set of real images) drawn from some unknown distribution $P_{\mathrm{data}}(x)$ generated by some unknown or ill-understood process, the objective here is to generate a new set (of images) whose probability distribution should match $P_{\mathrm{data}}(x)$ as closely as possible. During the training of the two models, the Generator samples a latent variable \(z\) from a known prior distribution \(P_z(z)\) ({\it e.g.}, the Normal Distribution) and produces a synthetic sample \(G(z)\) to start with and, subsequently, based on updates received from the Discriminator as its training progresses. The Discriminator, being a probabilistic binary classifier, receives either a real data sample \(x\) or a generated sample \(G(z)\), and outputs a probability that the input is real. The Discriminator aims to maximise its classification accuracy, while the Generator aims to fool it by trying to minimize it. The training of these two networks proceeds alternately leading to the optimization of the adversarial loss function $L(D,G)$ given by 
\begin{equation}
	\centering
	\begin{aligned}
		L(D,G) & = \min_{G} \max_{D} V(D, G)& \\
		&= \mathbb{E}_{x \sim p_{\rm data}(x)} \left[ \log D(x) \right] \\
		&+ \mathbb{E}_{z \sim p_{z}(z)} \left[ \log \left( 1-D(G(z)) \right) \right].
	\end{aligned}
	\label{eq:Basic_GAN}
\end{equation}
The two neural networks \( G(z) \equiv G(z; \theta_G)\) and \(D(x) \equiv D(x; \theta_D)\) are parameterized by \(\theta_i\) with \(i = G \ \mathrm{or}\ D\) respectively. During its training, the Generator samples random noise vector $z$ from a Normal Distribution and generates the differentiable function ${\hat x} \equiv G(z)$ that maps \(z\) to the data space \(x\). Through such maps the Generator generates its own distribution $p_{\mathrm G}({\hat x})$ and through the training episodes, specifically by iteratively updating its parameters $\theta_{\mathrm G}$, aligns this distribution with the distribution of real data $p_{\mathrm data}(x)$. The training data set provided to the Discriminator is constructed by mixing real data points $x$ and generated data points ${\hat x}$ in equal ratio. The Discriminator generates the function \(D(x)\) that represents the probability that \(x\) originates from real data. Eq.(\ref{eq:Basic_GAN}) implies that training of the GAN model moves towards maximization of the expectation of \(D(x)\). Through this process, both the parameters $\theta_{\mathrm G}$ and $\theta_{\mathrm D}$ are optimized such that $p_{\mathrm G}({\hat x})$ gets maximally aligned with $p_{\mathrm data}(x)$.  

For a given fixed Generator $G(z)$, the problem can be reformulated as:
\begin{equation}
	\centering
	\begin{aligned}
		\max_{D} V(D,G) &= \mathbb{E}_{x \sim p_{\rm data}} \left[ \log D^{*}_{G}(x) \right] \\ 
		&+ \mathbb{E}_{x \sim p_{G}} \left[ \log \left( 1 - D^{*}_{G}(x) \right) \right]
	\end{aligned}
	\label{eq:GAN_reformulated}
\end{equation}
where \(D^{*}_{G}\) denotes the optimum of the discriminator. As seen in equation (\ref{eq:Disc_optimum}), the global optimum of equation (\ref{eq:GAN_reformulated}) is achieved if and only if \(p_G = p_{\rm data}\). Furthermore, if both \(G\) and \(D\) are allowed to reach their respective optima, -- the so called Nash point of the Minimax game -- \(p_{\mathrm G}\) converges to \(p_{\rm data}\). 
\begin{equation}
	\centering
	D^*_G(x) = \frac{p_{\rm data}(x)}{p_{\rm data}(x) + p_G(x_{\mathrm gen})}
	\label{eq:Disc_optimum}
\end{equation}
At this point, the Discriminator finds its job no better than random guessing. A more comprehensive discussion of the problem, including proofs, is provided in \cite{goodfellow2014generative}.

Subsequently, the GAN framework was extended to a conditional model \citep{mirza2014conditional}. This new model, known as ``conditional GAN" or cGAN injects a conditioning variable \(y\) into both networks: the Generator now generates \(G(z \mid y)\), and the Discriminator evaluates \(D(x \mid y)\). The adversarial objective becomes  
\begin{equation}
	\centering
	\begin{aligned}
		V(D, G) &= \mathbb{E}_{x,y \sim p_{\rm data}(x)} \left[ \log D(x|y) \right] \\
		&+ \mathbb{E}_{z,y\sim p_{z}(z)} \left[ \log \left( 1-D(G(z|y)|y )\right) \right]
	\end{aligned}
	\label{eq:conditional_GAN}
\end{equation}
The conditional variable $y$ in a cGAN can be various additional information including images, labels or text contextual to ``ground truth'' real data $x$. Among various types of cGANs, Pix2Pix GAN with its image-to-image translation design is suitable for the task at hand.
In our work, we specifically use this conditional variable by choosing \(y\) to represent the ground-based intensity-interferometry (II) observation patterns: the Generator is trained to produce stellar surface images that not only look realistic but also confirm to the measured II correlations, while the Discriminator judges realism \emph{and} consistency with the II data.

\cite{isola2017image} further observed that combining the cGAN from Eq.~\eqref{eq:conditional_GAN} with the traditional $L_1(G)$ loss (also known  as the Mean Absolute Error) improves the results, as the Generator is encouraged to produce outputs closer to the target image in a pixel-wise sense. We adopt this approach in the training of our cGAN model by including $L_1(G)$ defined in eq.(\ref{eq:l1_loss})
\begin{equation}
	\centering
	L_1(G) = \mathbb{E}_{x \sim p_{\mathrm{data}, z \sim p_z(z)}} \bigl[ \lVert x - G(z \mid y) \rVert_{1} \bigr].
	\label{eq:l1_loss}
\end{equation}
The total adversarial loss function, along with the $L_1$ loss modulated by a hyperparameter $\lambda$ then becomes
\begin{equation}
	\centering
	L_{tot} = \arg \min_{G} \max_{D} V(D, G) + \lambda \cdot L_1(G).
	\label{eq:total_loss}
\end{equation}

This type of network has demonstrated remarkable robustness across a variety of applications. For example, it can generate colored images from grayscale inputs based on architectural labels, transform images from day to night, and even predict maps from satellite data. A more extensive list of applications is provided in \cite{isola2017image}.

\subsection{Generator}
As discussed above, in a GAN the Generator, a deep neural network in iteself, is responsible for producing synthetic data, in this case, images that resemble those of a fast-rotating star. In this work, the Generator is implemented as a U-Net convolutional network \citep{ronneberger2015u}. The U-Net consists of a symmetric encoder--decoder structure forming a characteristic ``U'' shape along with skip connections: the left (contracting or down-sampling) path, the right (expanding or up-sampling path) and the connecting (bottle-neck) path. The left down-sampling path repeatedly applies 3$\times$3 convolutions (padded to preserve spatial dimensions) followed by LeakyReLU activations and strided convolution (with stride of 2), by progressively halving the spatial resolution while doubling the channel depth (typically 64 $\to$ 128 $\to$ 256 $\to$ 512 $\to$ 1024). At the bottleneck, high-level features are processed without further spatial reduction. The right up-sampling path mirrors this process using 2$\times$2 transposed convolutions (stride 2) for up-sampling, halving the channel count at each level and using ReLU as the activation function for all its layers except the output layer. Additionally, a dropout layer is introduced at the beginning of the upsampling phase to mitigate overfitting of the Generator model \citep{isola2017image}. Critically, long skip connections concatenate feature maps from each encoder (down-sampling) level to the corresponding decoder (up-sampling level), directly injecting high-resolution details into the reconstruction process. This enables the network to ``remember where everything is'' while the deep bottleneck still provides the large receptive field needed to think globally about image structure and semantics. Besides the strided convolutions, modern variants of U-Nets used in state-of-the-art GANs incorporate residual blocks within resolution levels, and frequently add spectral normalization and self-attention at the bottleneck for improved training stability and long-range dependency modelling. These architectural choices allow our U-Net generator to simultaneously recover sharp, high-frequency details (stellar limb edges, limb-darkening profiles, gravity darkening, and rapid-rotation-induced oblateness) and enforce global coherence (overall disk morphology and physical consistency with the observed interferometric visibilities), making it ideally suited for high-fidelity sky-image reconstruction of fast-rotating stars from sparse ground-based intensity interferometry observations.

During the training, the total Generator Loss function $L_G$ including the $L_1$ loss defined in eq.(\ref{eq:l1_loss}) that is minimized is given by
\begin{equation}
\centering
\begin{aligned}
L_{G} \; &= \; - \, \mathbb{E}_{z \sim p_{z}(z)} \bigl[ \log D(G(z \mid y)\mid y) \bigr] \\
& +\; \lambda \; \mathbb{E}_{x \sim p_{\mathrm{data}, z \sim p_z(z)}} \bigl[ \lVert x - G(z \mid y) \rVert_{1} \bigr].  
\end{aligned}
\label{eq:total_gen_loss}
\end{equation}
Here, 
\begin{itemize}
  \item \(x\) denotes a real data sample (e.g.\ the ``ground-truth'' image; here, the synthetically generated fast rotator image) corresponding to condition \(y\), the simulated II observation data.  
  \item \(z\) is a random latent vector drawn from the prior \(p_z(z)\).  
  \item \(G(z \mid y)\) is the generator output (the reconstructed / synthesized image) given \(z\) and condition \(y\).  
  \item \(D(\cdot \mid y)\) is the discriminator’s estimate (probability) that its input is “real,” given the same condition \(y\).
  \item \(\lambda\) is a hyperparameter controlling the trade-off between adversarial realism and pixel-wise fidelity (typical values depend on the problem, e.g.\ in pix2pix, \(\lambda = 100\)). 
\end{itemize}

\subsection{Discriminator}
The Discriminator is tasked with classifying the images produced by the Generator as either real or fake. It takes a real image from the dataset (often referred to as the target image for the Generator) and provides feedback to guide the Generator toward producing more accurate images. In this work, the PatchGAN model \citep{isola2017image} is employed as the Discriminator. Unlike a traditional global classifier, PatchGAN evaluates individual patches of the image, outputting a grid of predictions rather than a single scalar value. Each element in the grid corresponds to the ``realness'' of one patch of the image under examination of the Discriminator at a time. The final loss of the Discriminator is the average over all the patch responses. Evaluating the ``realness'' of the input image in terms of its constituent patches facilitates capture of texture/ style and other high frequency components in the image. As compared to a global discriminator, it also reduces the number of parameters in the network thereby helping reduce computation cost. It also works on images with arbitrary sizes.

Prior to the down-sampling of data using PatchGAN, each input image is preprocessed with application of Zero Padding followed by batch normalization. The purpose of Zero Padding is to prevent the loss of spatial information and to facilitate the extraction of deeper features from the down-sampled output. Batch normalization is required to stabilize the learning (loss minimization) process. PatchGAN then reduces the spatial dimensions of the images to extract localized features, ensuring the model focuses on smaller regions. In this down-sampling stage, a leaky version of the Rectified Linear Unit (LeakyReLU) is applied in the convolutional layers, similar to the approach used in the Generator. The probability $D(\cdot|y)$ that the patch of the input image represented by $\cdot$ is ``real'' is evaluated through this process. The loss function $D$ of the full input image is obtained by averaging over all the patch responses. 

The Discriminator Loss function $L_D$ that is optimized during the training process is given by 
\begin{equation}
\centering
\begin{aligned}
L_{D} \; &= \; - \, \mathbb{E}_{x \sim p_{\mathrm{data}}(x \mid y)} \bigl[ \log D(x \mid y) \bigr] \\
& -\; \mathbb{E}_{z \sim p_{z}(z)} \bigl[ \log\bigl( 1 - D(G(z \mid y)\mid y)\bigr) \bigr]
\end{aligned}
\end{equation}
where the arguments of the $D$ and $G$ functions are as noted in the text following eq.(\ref{eq:total_gen_loss}) This loss is composed of two parts: one that assesses how accurately it identifies fake images (by comparing predictions to a target value of 0) and the other that measures how accurately the Discriminator identifies real images (by comparing predictions to a target value of 1).

The training procedure of these two components of the cGAN model can be outlined as follows:
\begin{itemize}
\item{The discriminator \(D\) is updated by minimizing \(L_D\), keeping \(G\) fixed (so that \(D\) becomes better at distinguishing real vs generated images).}  
\item{The generator \(G\) is updated by minimizing \(L_G\), keeping \(D\) fixed, thus pushing \(G\) to generate images that are (a) judged “real” by \(D\), and (b) close (in pixel-wise sense) to the ground-truth \(x\).} 
\end{itemize}

\section{Network Parameters}

\begin{figure*}
	\centering
	\includegraphics[width=0.97\textwidth]{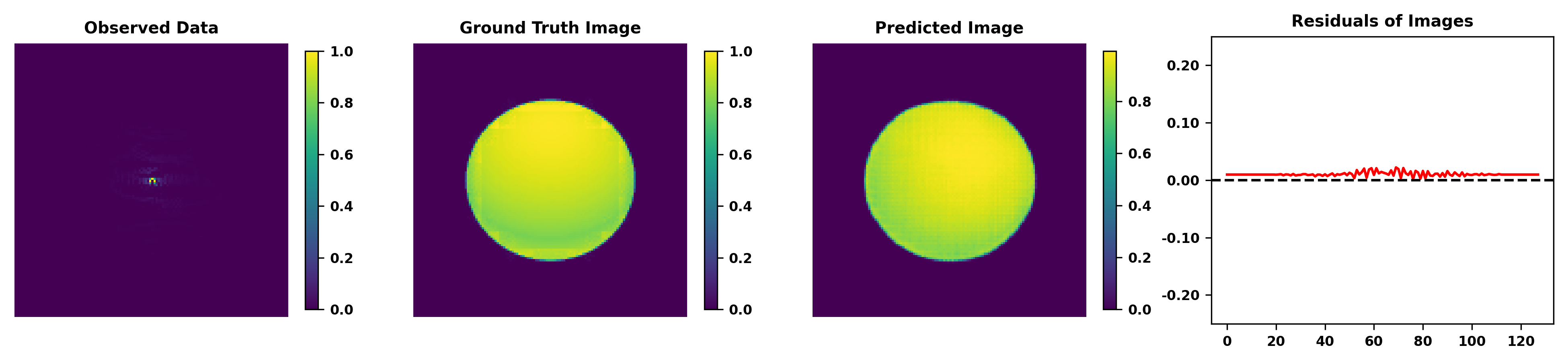}
	\includegraphics[width=0.97\textwidth]{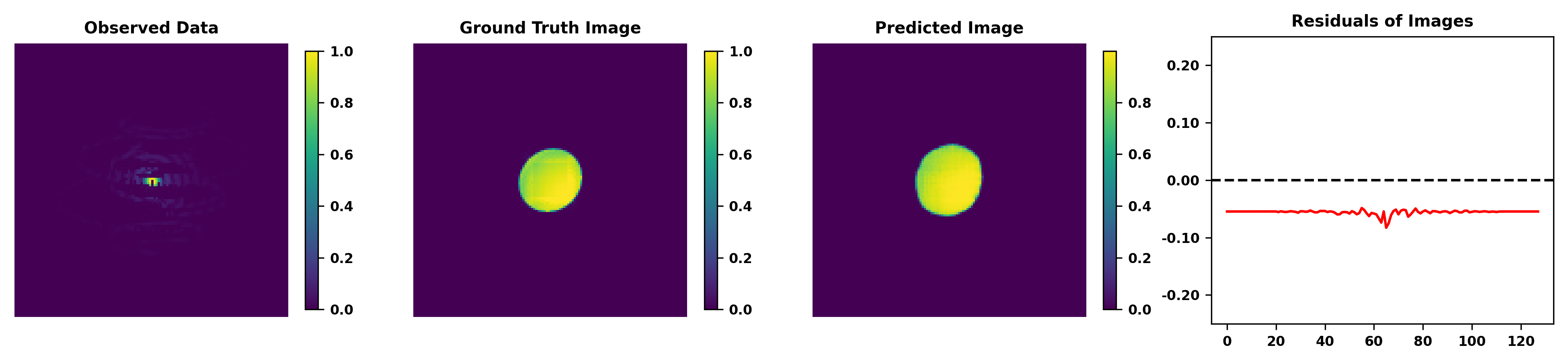}
	\includegraphics[width=0.97\textwidth]{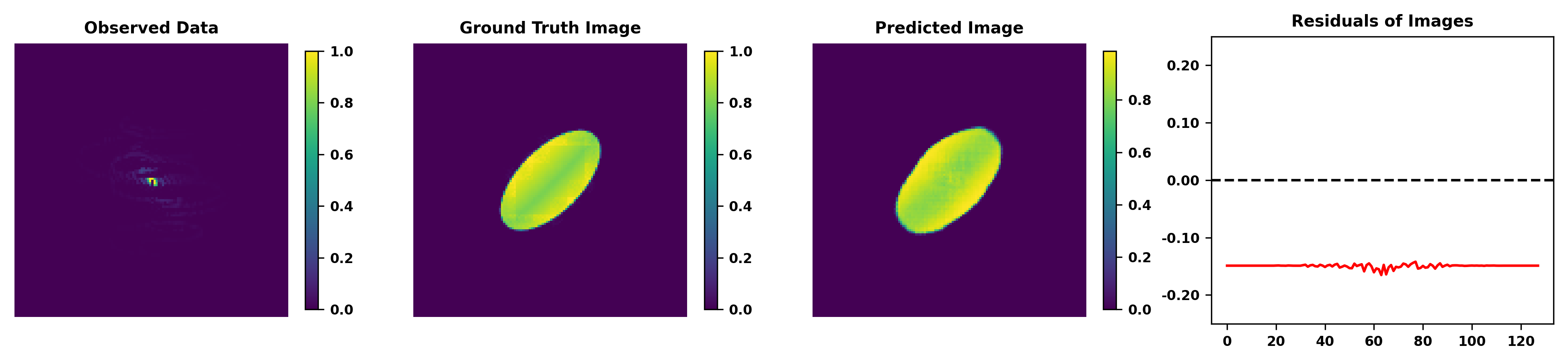}
	\includegraphics[width=0.97\textwidth]{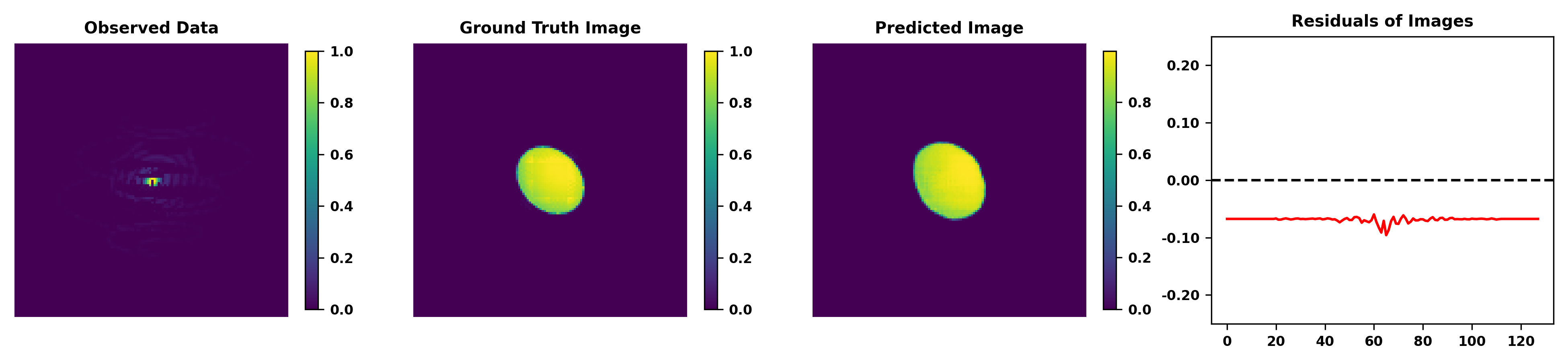}
	\includegraphics[width=0.97\textwidth]{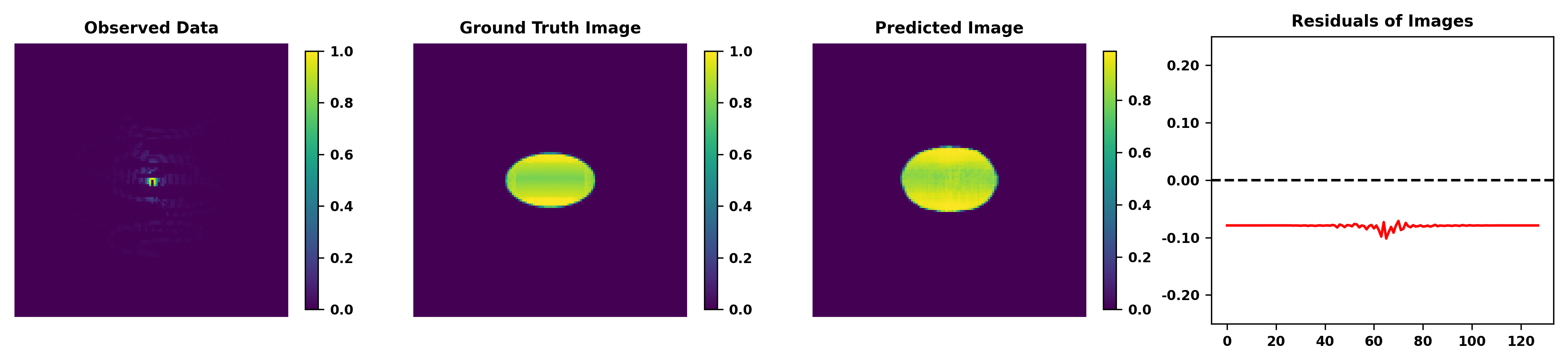}
\caption{Example results of image reconstruction using the cGAN model along with the II observations simulated in this work. Each row in this figure represents the results for a hypothetical fast-rotating star. The left panel represents the sparse II pattern obtained by the simulated observation of the star using the array of $N_T =6$ IACTs illustrated in the left panel of Fig.\ref{fig:teles}. This image acts as the ``condition'' part of the data input to the cGAN model. The second panel from left displays the real image, or ground truth, which the Discriminator uses to distinguish from the images generated by the Generator. The data generated for training, validation and testing of the cGAN model is a merge of this image and its II pattern presented in the left panel. The third panel is the reconstructed image, or the predicted image, produced by the trained GAN model . The fourth and the last panel shows the 1D-residuals between the ground truth and the predicted image in the $(u,v)$ plane. The x-axis of this panel represents the horizontal pixel coordinate and the y-axis shows the residual intensity value at the respective pixel.}
	\label{fig:six_GAN}
\end{figure*}

\begin{figure*}
	\centering
	\includegraphics[width=\textwidth]{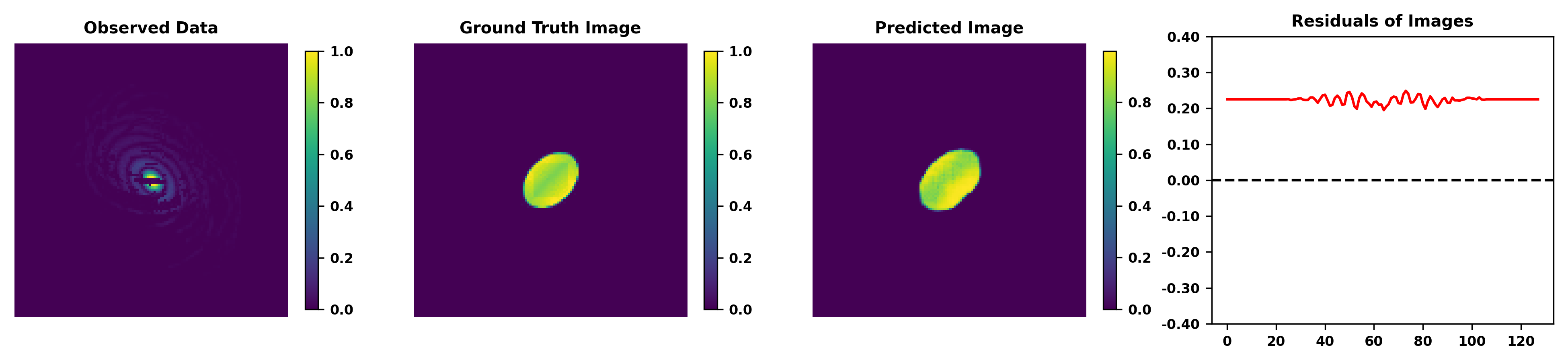}
	\includegraphics[width=\textwidth]{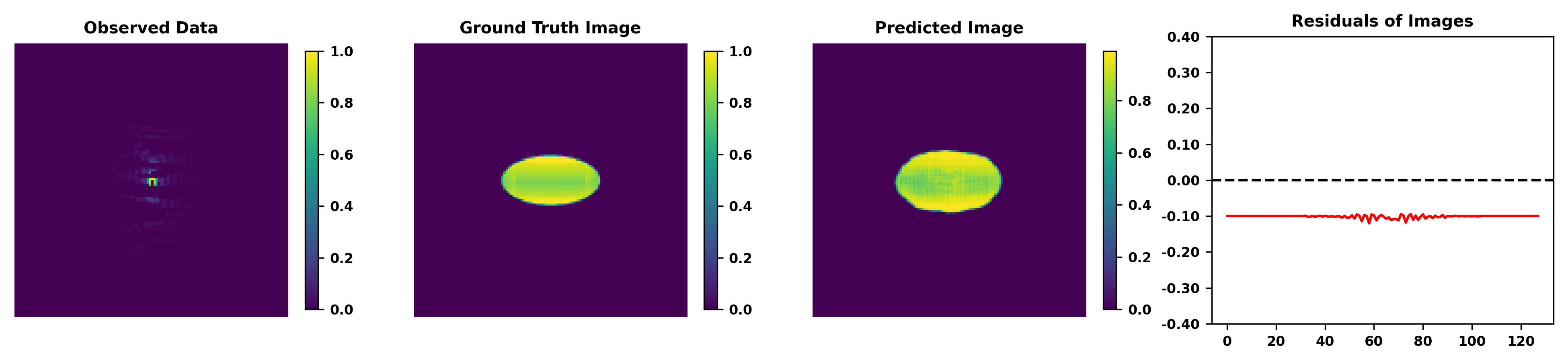}
	\includegraphics[width=\textwidth]{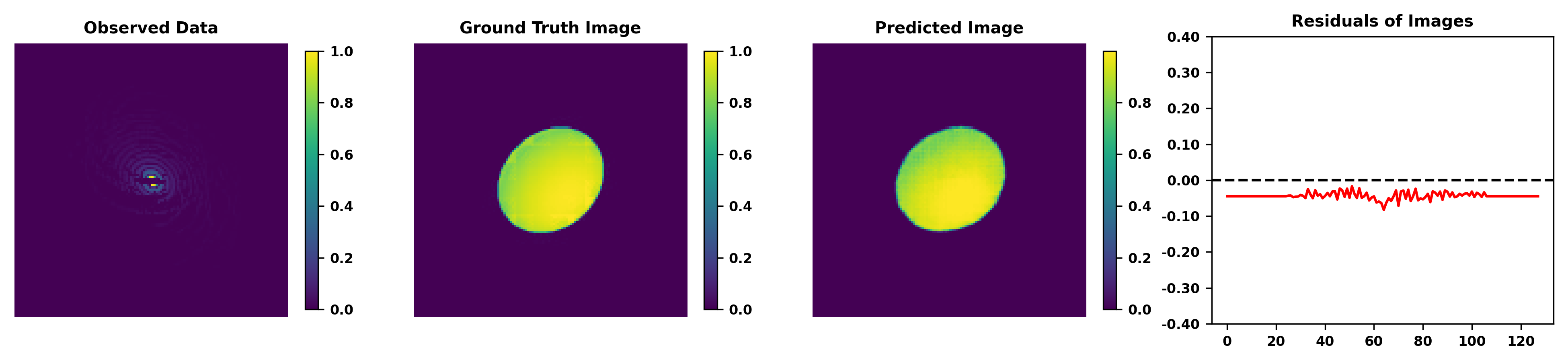}
	\includegraphics[width=\textwidth]{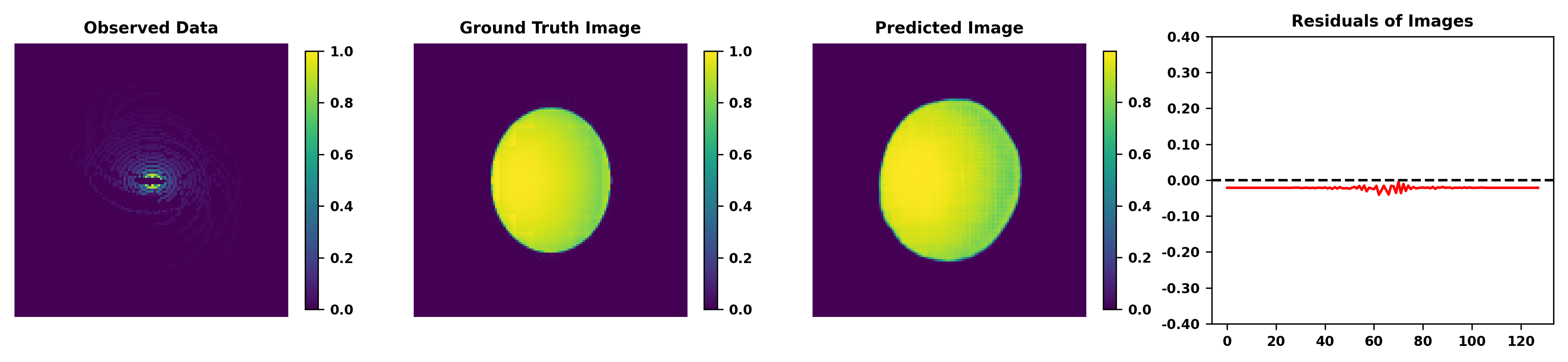}
	\includegraphics[width=\textwidth]{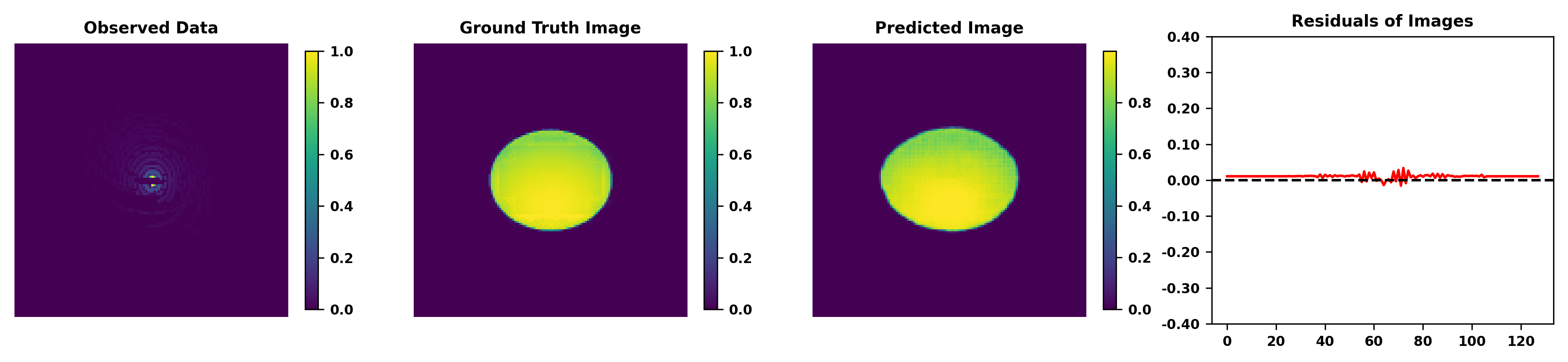}
	\caption{Example results of image reconstruction using the cGAN model (as presented in Fig.~\ref{fig:six_GAN}), but with the II simulations carried out with the array of $N_T =9$ as shown in the right panel of Fig.~\ref{fig:teles}.  Panel labels are as described in the caption of Fig.~\ref{fig:six_GAN}.} 
	\label{fig:nine_GAN}
\end{figure*}

An appropriate cGAN architecture along with a set of hyperparameters was optimized by tuning. The objective of the model, as already mentioned, was to learn to faithfully reproduce a set of sky-images of fast rotators subject to the condition that those images are consistent with their simulated II observation data. We discuss below the architecture and the hyperparameters of the cGAN model used for this task. Given the adversarial nature of GANs, where the Generator and Discriminator engage in a Minimax game, careful tuning of key parameters is critical to ensure that both networks are well-balanced for effective training.

\subsection{Data Preparation}\label{sec:DataPrep}
First, we generate synthetic images of rapidly rotating stars by modelling them as oblate spheroids with varying radii and oblateness parameters between 0.5 and 1. To incorporate the effect of gravity darkening, we also consider different viewing angles and assume a linear dependence on the declination angle of each point on the stellar surface. The traced ellipses result from integrating over the source's hour angle.

Next, Salt and Pepper noise is introduced into these images; usually this is done at the rate of 0.5\% of the number of pixels in the image. Then, the images are resized and their mean is subtracted. A two-dimensional Fast Fourier Transform, along with a Fourier shift, is applied, yielding a complex number for each pixel. Since II does not measure phase, the absolute value is calculated. 

Next, sampling of the interferometric plane is introduced via pixel-wise multiplication of the absolute-valued Fourier-transformed image (left panel of Fig.~\ref{fig:ft}) and the two sets of baseline tracks of (Fig.~\ref{fig:base}). It may be recalled that these two sets of baseline tracks of Fig.~\ref{fig:base} are, respectively, generated by each of the distinct pairs of telescopes in each of the two IACT arrays of Fig.~\ref{fig:teles}; i.e., the array of six IACTs (the left panel) and the array of nine IACTs (the right panel). These tracks are generated due to the Earth's rotation during the period (one night) of II observation. The result of this pixel-wise multiplication is a set of two sparsely sampled maps in the Fourier plane featuring several ellipses and are shown in the middle and the right panels of Fig.~\ref{fig:ft}. In essence, these two panels of fig.~\ref{fig:ft} represent the results of simulated II observation of the fictitious fast-rotator illustrated in fig.~\ref{fig:image} carried out respectively by the two IACT arrays depicted in Fig.~\ref{fig:teles}.

Finally, we normalize the pixel values and convert them to 8-bit integers, producing an image that encodes the sparsely sampled, phase-free visibility measured by II. The image of the corresponding simulated (fictitious) star, which serves as the ``ground truth'' is processed identically to avoid any bias. These two images are merged side-by-side into a single 
image (as shown in Fig.~\ref{fig:GANinput}). Thus two pairs of datasets, each corresponding to one of the two IACT arrays, are created. Each of these datasets are then split into three parts in the proportions of 80\% for training, 10\% for validation, and 10\% as the test set. This partition of the full dataset is indicated in the Fig.\ref{fig:FlowDiagram}. In the following, we refer to these three parts as the Training Set, the Validation Set and the Testing Set respectively.

\subsection{GAN Architecture}
The GAN architecture used in this work is a Pix2Pix cGAN, which uses an image-to-image translation strategy with both the ground truth and the condition being images. Originally introduced by Isola et al. \citep{isola2017image}, this architecture is specifically suitable for image processing and reconstruction objectives. For instance, the TensorFlow tutorials\footnote{\url{https://www.tensorflow.org/tutorials/generative/pix2pix}} demonstrate its application to a dataset of architectural facades. This architecture has been adapted
for the phase retrieval problem at hand here. The network is implemented using the TensorFlow library \citep{abadi2016tensorflow}, calculations are performed with scipy \citep{virtanen2020scipy}, and plots are generated with matplotlib \citep{4160265}.

\subsection{Hyperparameter Tuning}
The cGAN model architecture used in this work employs several hyper-parameters, which are explained briefly below \citep[for a more in-depth discussion, see][]{murphy2022probabilistic}.

The learning rate ($lr$) of the optimizer determines how much the model updates its parameters with each iteration. A learning rate that is too small may lead to underfitting, while one that is too large can render the model unstable. Therefore, selecting an appropriate learning rate is crucial \citep{murphy2022probabilistic}. A canonical choice in Pix2Pix and other GAN models is $lr = 2\times10^{-4}$. In our case too, we found this choice to be appropriate.

The kernel size refers to the dimensions of the convolutional kernel used in the network, determining how many pixels are combined to produce a new pixel. A larger kernel size can capture features spanning several pixels, but it may also incorporate unrelated features. Small kernel sizes are preferable in cases where target images contain finer details or high spatial frequency features. Since the ``ground truth'' target images in our case have longer scale gravity darkening features, we have opted for the more canonical choice of kernel size being 5.

The amount of noise is controlled by two parameters, ``alpha" and ``beta", which indicate the percentage of pixels altered to either white or black, hence the term Salt and Pepper noise. Here, ``alpha" is applied to the real image, while ``beta" is applied to the generated image. Different ratios (``alpha/beta") can lead to varying model performance; however, our results indicate that distinct noise rates do not significantly affect the loss functions. 

The batch size defines the number of images processed simultaneously by the network. Smaller batch sizes have been observed to improve generalization \citep{prince2023understanding}. However, because a larger batch size significantly increases training time, a batch size of 1 is used. Besides, in Pix2Pix cGAN implementations found in literature, this choice is found to be often preferred.

The buffer size of a Pix2Pix GAN refers to a small memory of image pool of previously generated images. They are occasionally fed to the Discriminator in place of the freshly generated ones. This strategy of mixing old and new fakes  mitigates the risk of mode collapse wherein the Discriminator tends to map all or large number of generated images to only one or a few real ``ground truth'' image(s). We have chosen a fairly large buffer size (=1400) to protect the model against mode collapse.

In the training of GANs, one often-followed strategy to potentially boost performance is to give the Discriminator an advantage by increasing its number of training epochs before returning to the Generator's training. This hyperparameter is referred to as the Discriminator repetition (as seen in Table \ref{tab:hyperparameters}). While this can lower the Discriminator loss, as shown in Fig.~\ref{fig:Plot_discrep_loss}, it also increases training time. In training our model, we did not notice any significant advantage derived from this strategy. Since the generated images did not noticeably improve with additional Discriminator training, we adopted the strategy of training both the networks with equal preference (Discriminator repetition = 1).

One domain specific hyperparameter of the cGAN model presented here is the Number of Telescopes $N_T$. The degree of sparse sampling of the intensity interferometric (II) image plane can be varied to provide the model with access to more number of active (non-zero) pixels. Point to note here, is that the coverage of the Fourier interferometric plane (number of active pixels in the II image) scales with available number of baselines, and the latter scales quadratically with the number of telescopes.

The hyperparameter Output Channels refers to the number of channels in the generator output (e.g., 1 for grayscale, 3 for RGB). It is worthwhile to recall that the cGAN model constructed in this work is trained on ``ground truth'' target images and the simulated II data, both in grey scales as seen in Fig.\ref{fig:GANinput}. It is natural that the output of this model will be in grey scales only. Therefore the value of this hyperparameter is set to 1. The choice of hyperparameter \(\lambda\) has been commented upon earlier.

An optimized set of hyperparameters is selected through an iterative process of training and validation. For a tentatively chosen set of the hyperparameters, the model is trained using the Training Set until the both the Discriminator and the Generator loss functions are minimized. This model, thus trained along with its model parameters, is then passed through validation using the Validation Set. This cycle of training and validation is iterated till an optimal set of hypermaraters is arrived at. During each epoch of training of the model, both the Discriminator and the Generator networks are trained for 100,000 steps. Plots of the ``Discriminator Loss'' function and the ``Generator Total Loss'' function presented in Fig.{\ref{fig:Plot_discrep_loss}} represent the results of this training for the hyperparameters, namely, the Disriminator repetition (``discrep'', in short). Obviously, the most compute-intensive part of this process is that of the training of the Model. The results of training and validation presented in this work were carried out on a CPU using two nodes, each with 48 threads and the entire process of training and validation required approximately 20 hours on the machine employed for this work. This iterative process of training and validation of the Model is represented schematically in the Fig.\ref{fig:FlowDiagram}. The chosen optimal set of hyperparameters is presented in Table-{\ref{tab:hyperparameters}.}
\begin{table}[ht]
	\centering
	\caption{Selected hyperparameters used for training the model.}
	\label{tab:hyperparameters}
	\begin{tabular}{ll}
		\hline
		\textbf{Hyperparameter} & \textbf{Value} \\
		\hline
		Learning rate           & 2e-4 \\
		Kernel size             & 5 \\
		Alpha/Beta              & 1 \\
		Batch size              & 1 \\
		Buffer Size             & 1400 \\
		Discriminator repetition  & 1 \\
		Number of Telescope        & 4 \\
		Output Channels         & 1 \\
		Lambda                  & 100 \\
		\hline
	\end{tabular}
\end{table}
This optimized and trained Model is then subjected to testing and evaluation using the Testing Set. The results of this testing and evaluation is presented in the following section.

The Pix2Pix cGAN architecture along with the choice of the values of the hyperparameters mentioned in the Table -\ref{tab:hyperparameters} and used in the implementation of this architecture constitutes the cGAN model (hereinafter referred to as ``the GAN Model'' or simply ``the Model'').

\section{Image Reconstruction: Results and Evaluation}
\begin{figure*}
	\centering
	\includegraphics[width=.44\linewidth]{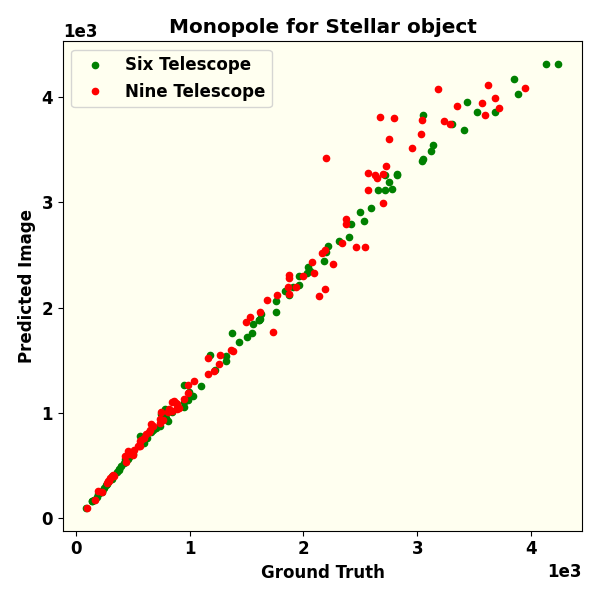}
	\includegraphics[width=.44\linewidth]{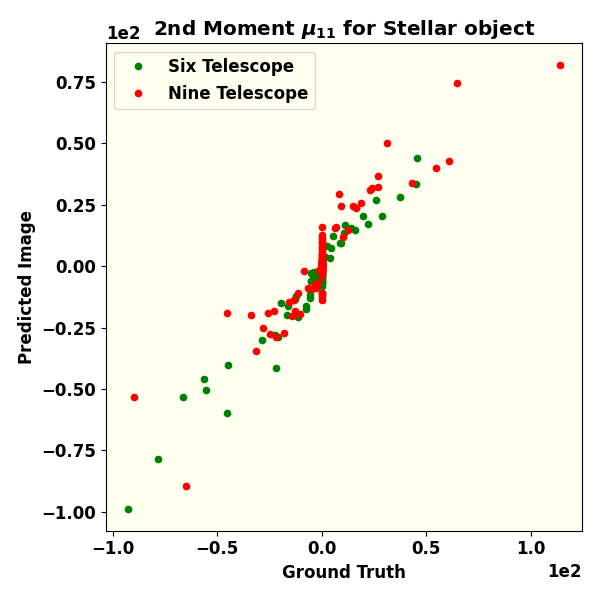}\hfill
	\includegraphics[width=.44\linewidth]{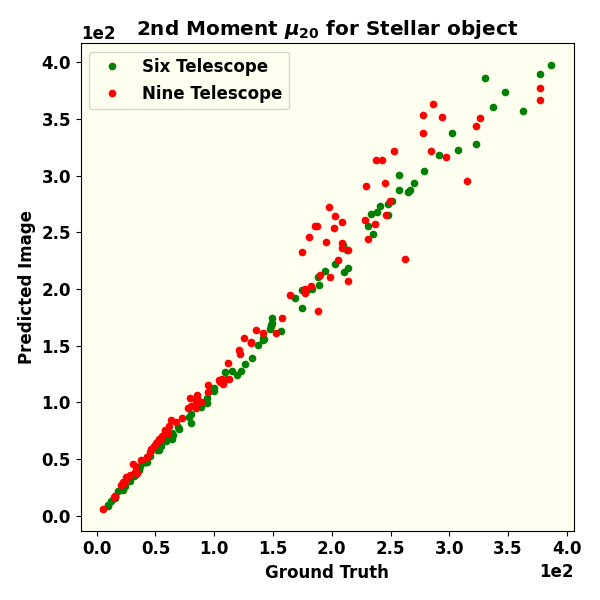}
	\includegraphics[width=.44\linewidth]{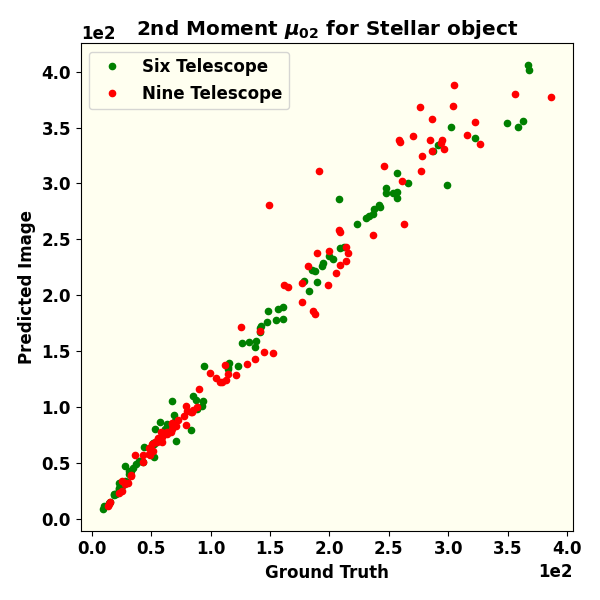}\hfill
	\includegraphics[width=.44\linewidth]{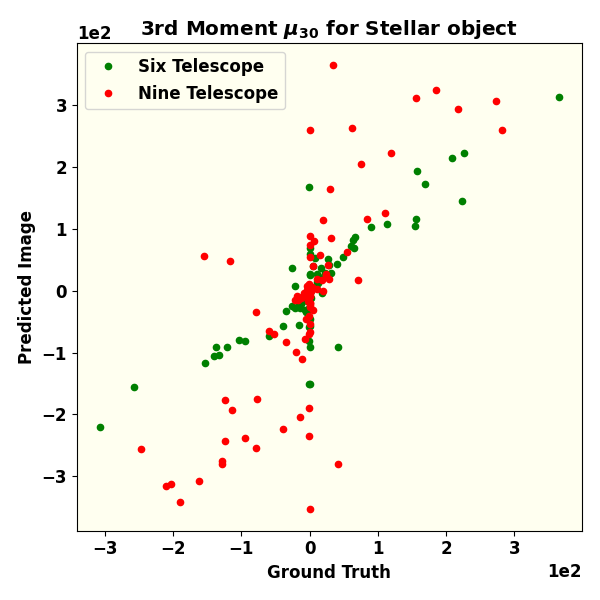}
	\includegraphics[width=.44\linewidth]{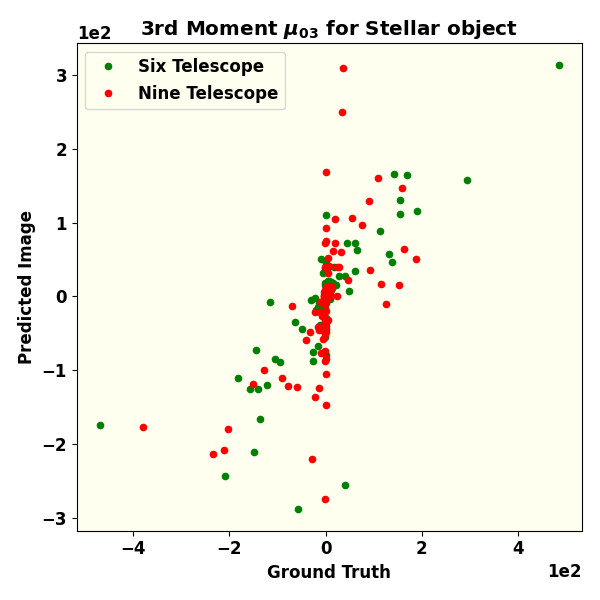}\hfill
	\caption{Continue figures on next page.}
	\label{fig:struc}
\end{figure*}

\begin{figure*}\ContinuedFloat
	\centering
	\includegraphics[width=.44\linewidth]{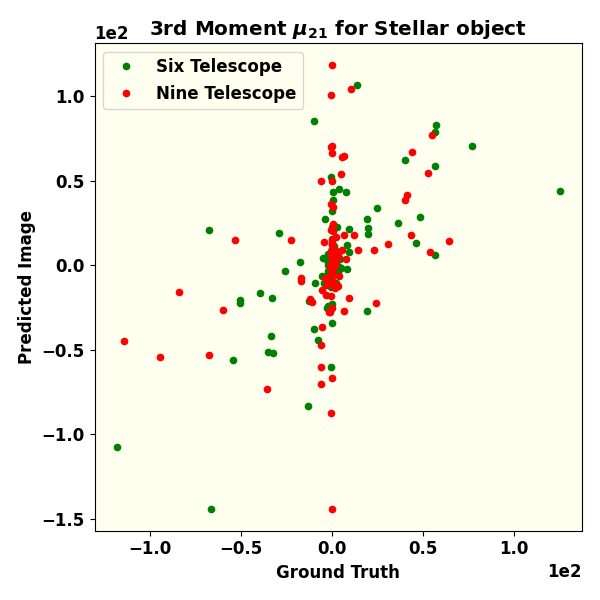}
	\includegraphics[width=.44\linewidth]{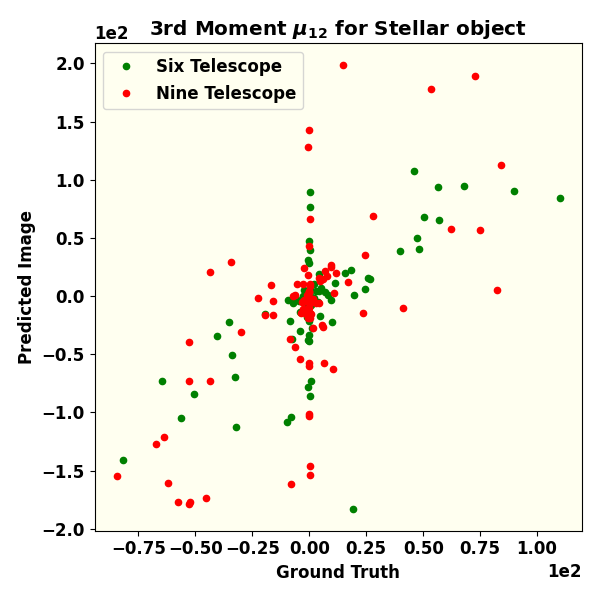}	\hfill
	\caption{The panels of this figure depict the comparison of moments of predicted images vs. those of the ground truth images. The predicted images are generated by our trained cGAN model vis. a vis. the ground truth images in the two Test Sets. The green and the red coloured dots represent the moments of the FRSs in these Test Sets generated with II observations with the arrays of $N_T=6$ and of $N_T=9$ IACTs. The small scatter in the moments of the predicted images is indicative of a balanced training of the Model. This figure also shows that improvement in (u, v) coverage dominate over the number of baselines.}
\end{figure*}                     
In this section, we examine the performance of the GAN Model whose architecture and choice of hyperparameters have been discussed above. We subject the trained Model to the task of phase retrieval and image reconstruction on the Testing Set. In this process, for each ground truth image of the Testing Set, two predicted images are generated: one each for the arrays of $N_T=6$ and $N_T=9$ IACTs. 

\subsection{Visual Evaluation of Images Predicted by the Model}
Fig.~\ref{fig:six_GAN} and Fig.~\ref{fig:nine_GAN} demonstrate the success of the trained Model in reconstructing the images of a sample of the fictitious fast-rotators drawn from the Testing Set.

The four panels from left to right in each row of Fig.~\ref{fig:six_GAN} and Fig.~\ref{fig:nine_GAN} represent the following:
\begin{itemize}
\item{The left panel represents the sparse II pattern obtained by the simulated observation using the set of IACTs illustrated in Fig.\ref{fig:teles}. As explained earlier, this image acts as part of the input, namely its ``condition'' part. This image acts as the condition that the sky-images of the star generated by the Generator must conform to.}
 
\item{The second panel from left displays the real image, or ground truth target image, which the Discriminator loss function uses to distinguish from the images generated by the Generator.}

\item{The third panel from left presents the reconstructed (or predicted) image corresponding to the ground truth (second panel) and generated by the trained Model. The similarity of these two images indicates the success of the Model in its stated objective of image reconstruction. We remark that rotating the images by $180^\circ$ around the rotation axis of the star would not change the image.  That is, each predicted image contains an arbitrary choice among two possible orientations, differing by $180^\circ$.}

\item{The right panel shows the 1D residuals between the ground truth target image and the predicted image in the interferometric plane. The small values are indicative of the success of training the Model.}
\end{itemize}

The predicted images and the residuals presented in the third and the fourth columns (from left) of Fig.~\ref{fig:six_GAN} and Fig.~\ref{fig:nine_GAN} show encouraging results, conveying visual information about the source's size, shape, and brightness distribution across its surface fairly accurately. This has been achieved on the basis of the input provided by II observation using only 15 and 36 baselines, corresponding to six and nine telescopes, respectively.  Further improvements could surely be achieved by increasing the number of telescopes to enhance the coverage of the $(u, v)$ plane. A closer examination of this proposition might be able to contribute to the design and instrumentation aspects in the existing and upcoming CTAO.

\subsection{Evaluation of the Model using Moments}
The reconstructed images are visually compelling, demonstrating the Model's effectiveness in using II to reconstruct images. However, visual assessment alone is insufficient; statistical evaluation of the generated images in comparison with the ground truth images is necessary to validate the results. We present here the results of our calculation and comparison of the moments of distribution of pixel brightness in the target ``Ground Truth'' images and the ``Predicted'' images.

Image moments capture key properties of the reconstructed objects, such as shape, size, and intensity distribution, by quantifying features like position, orientation, and brightness distribution. By comparing the moments of the Model-generated images with those of the ground truth, we can objectively assess the consistency and accuracy of the reconstruction. This approach provides a reliable framework for evaluating reconstruction quality, as image moments can reveal subtle differences in geometric and intensity properties that might not be apparent through visual inspection alone.

The raw moment $M_{ij}$ of an image is defined as \citep{hu1962visual}
\begin{equation}
	M_{ij} = \sum_{x} \sum_{y} x^i y^j I(x, y)
	\label{eqn:Mom-ij}
\end{equation}
where $I(x,y)$ represents the intensity at pixel $(x,y)$. The zeroth order raw moment, or monopole, represents the total intensity of an image. It is computed by summing all pixel values across the image, yielding an overall intensity measure. In this context, analyzing the monopole provides the total pixel brightness of the images of the fictitious stars. According to Eq.~\eqref{eqn:Mom-ij}, the monopole of an image is calculated as:
\begin{equation}
	M_{00} = \sum_{x} \sum_{y} I(x, y).
	\label{eqn:Mom-00}
\end{equation}

The first order raw moments normalized by the respective monopole moments of the images given by 
\begin{eqnarray}
&&x_c = \frac{M_{10}}{M_{00}} = \frac{\sum_{x,y} x \cdot I(x,y)}{\sum_{x,y} I(x,y)} \nonumber \\
&&y_c = \frac{M_{01}}{M_{00}} = \frac{\sum_{x,y} y \cdot I(x,y)}{\sum_{x,y} I(x,y)}
\label{eqn:Centroid}
\end{eqnarray}
represent the centroids $(x_c, y_c)$ of the pixel brightness distribution of the images.

Furthermore, these calculated centroids are instrumental in analyzing the shape, size, and brightness distribution of the stars using higher-order image moments. To this end, the central moment of an image is calculated according to:
\begin{equation}
	\mu_{pq} = \frac{1}{M_{00}}\sum_{x} \sum_{y} (x - x_c)^p (y - y_c)^q I(x, y).
\label{eqn:Central-Mom}
\end{equation}
The sum of \(p\) and \(q\) defines the order of the central moment. 

The second order ($p+q\,=\,2$) central moment of the brightness distribution of the images is analogous to the moment of inertia of a mass distribution. 

Fig.~\ref{fig:struc} presents the comparison of monopole, second-order central moments (\(\mu_{11}, \mu_{20}, \mu_{02}\)), and third-order central moments (\(\mu_{30}, \mu_{03}, \mu_{21}, \mu_{12}\)) which are used to study the shape, structure and brightness distribution of a fast-rotating star along the line of sight, respectively using 6- and 9- IACT facilities (as explained in the upcoming subsection). The figure also highlights the impact of (u,v)-plane coverage on image reconstruction.

The green data points correspond to the monopole and moments obtained with six telescopes arranged as shown in the left panel of Fig.~\ref{fig:teles}, while the red points represent results from the nine-telescope configuration displayed in the right panel of the same figure. The red points exhibit a comparatively larger scatter than the green ones, suggesting that improved (u,v)-plane coverage plays a more significant role than simply increasing the number of baselines.

A numerical comparison of scatter of the moments of the reconstructed (or predicted) images with reference to the ground truth images is worked out and presented here. In order to evaluate the ``scatter metrics" of the moments, let us first extend the notations given in the formulas of raw moments (as in eq.(\ref{eqn:Mom-ij}), the centroids (as in eqs.(\ref{eqn:Centroid}) and the central moments (as in eq.(\ref{eqn:Central-Mom}) and apply them to the three sets of images produced in the testing process. Thus we define the raw moments $M_{ij}^{\mathrm{GT}}$, $M_{ij}^{\mathrm{N6}}$ and $M_{ij}^{\mathrm{N9}}$ for each image belonging to the set of ground truth  (``$\mathrm{GT}$"), and for the reconstructed ``$\mathrm{N6}$", ``$\mathrm{N9}$" image sets respectively. Extending the same logic, we use the eqs.(\ref{eqn:Centroid}) to define the centroids
$\left\{\left(x_c^{\mathrm{GT}}, y_c^{\mathrm{GT}}\right),
\left(x_c^{\mathrm{N6}}, y_c^{\mathrm{N6}}\right),
\left(x_c^{\mathrm{N9}}, y_c^{\mathrm{N9}}\right)\right\}$
respectively for each of the images in the three data sets.
Once again, using eqn.~(\ref{eqn:Central-Mom}), for each image in each of the datasets, we can define the central moments.

Now, the scatter of the centroids $S_c^\mathrm{N6}$ for the array of $N_T=6$ IACT is given by
\begin{equation}
S_c^{\mathrm{N6}} = \frac{1}{N}\sqrt{\sum_{i=1}^N\left[(x_{c,i}^{\mathrm{GT}} - x_{c,i}^{\mathrm{N6}})^2 +(y_{c,i}^{\mathrm{GT}} - y_{c,i}^{\mathrm{N6}})^2 \right]} 
\label{eqn:scatter-c}
\end{equation}    
where the summation goes over all the images in the two sets. Similarly, the scatter $S_{pq}^{\mathrm{N6}}$ of the central moments of the $\mathrm{N6}$ set of predicted images with reference to the ground truth images can be defined as
\begin{equation}
S_{pq}^{\mathrm{N6}} = \frac{1}{N}\sqrt{\sum_{i=1}^N \left(\mu_{(pq),i}^{\mathrm{GT}} - \mu_{(pq),i}^{\mathrm{N6}}\right)^2}
\label{eqn:scatter-CentMom}
\end{equation}
where the order parameter of the central moments is decided by the set of integer values $\{p,q = 0,1,2,3\}$ and the order of the central moment being the the sum of $p$ and $q$. The scatter metrics 
$S_c^\mathrm{N9}$ and $S_{pq}^{\mathrm{N9}}$ for the predicted images of the $\mathrm{N9}$ set can also be defined and calculated using formulas similar to eqs.(\ref{eqn:scatter-c}) and (\ref{eqn:scatter-CentMom}). Table~\ref{tab:mom} summarizes these scatter metrics between the ground-truth and reconstructed images for the centroids of the images and each central moment. The comparison clearly shows that the six-telescope configuration yields smaller scatter values, whereas the nine-telescope configuration results in larger deviations. This further supports the conclusion that optimized $(u,v)$-plane coverage is more critical than the total number of baselines. 

Overall, the close agreement between the ground truth and predicted moments demonstrates the effectiveness of the model in reconstructing images using intensity interferometry. The relatively small scatter in the predicted moments further indicates that the model has learned the underlying features robustly, without significant over-fitting.
\begin{table*}
	\centering
	\caption{Scatter metrics of the two predicted sets with reference to the set of ground-truth images (see Fig.~\ref{fig:struc}). Here, $S_c$ denotes the scatter of the centroid ($\mathrm{c}$) of the images, and $S_{pq}$ the scatter of the central moments of order $p+q$ where the indices $\{p,q = 0,1,2,3\}$).}
	\label{tab:mom}
	\begin{tabular}{ccccccccc}
		\toprule
		Predicted Set & $\mathrm{S}_\mathrm{c}$ & $\mathrm{S}_{11}$ & $\mathrm{S}_{20}$ & $\mathrm{S}_{02}$ & $\mathrm{S}_{12}$ & $\mathrm{S}_{21} $ & $\mathrm{S}_{30}$ & $\mathrm{S}_{03}$ \\
		\midrule
		$\mathrm{N6}$ & 0.0223 & 0.387 & 7.606 & 7.296 & 35.822 & 30.603 & 102.691 & 91.102 \\
		$\mathrm{N9}$ & 0.0256 & 0.499 & 11.446 & 10.869 & 36.159 & 35.865 & 108.123 & 108.622 \\
		\bottomrule
	\end{tabular}
\end{table*}
\subsection{The reconstructed Parameters for object}
The centroids \((x_c, y_c)\) indicate only the center of the star and its spatial location in the image. In contrast, the second-order central moments determine the orientation, semi-major axis, and eccentricity relative to the source's center \citep{teague1980image}. These moment-based parameters fully describe the two-dimensional ellipse that fits the image data.

The orientation of a fast-rotating star along the line of sight is defined in terms of second-order central moments as
\begin{equation}
	\theta = \tfrac{1}{2}\arctan \left(\frac{2\mu_{11}}{\mu_{20} - \mu_{02}}\right).
	\label{eqn:orn}
\end{equation}
The semi-major and semi-minor axes of the stellar object are computed using the second-order central moments and are denoted as \(a\) and \(b\), respectively.
\begin{equation}
	\begin{aligned}
		&a = 2\sqrt{mp + \delta} \\
		&b = 2\sqrt{mp - \delta}
	\end{aligned}
	\label{eqn:semi}
\end{equation}
where,
\begin{equation}
	mp = \frac{\mu_{20} + \mu_{02}}{2}
	\label{eqn:mp}
\end{equation}
and
\begin{equation}
	\delta = \frac{\sqrt{4\mu_{11}^2 + (\mu_{20} - \mu_{02})^2}}{2}.	
	\label{eqn:delta}
\end{equation}
Using the calculated axis values, the eccentricity of the fast-rotating star is determined as:
\begin{equation}
	e = \sqrt{1 - a/b}.
	\label{eqn:eccen}
\end{equation}
Eqs.~\ref{eqn:orn}-\ref{eqn:eccen} describe the elliptical nature of the stellar object (in this case, a fast-rotating star) and provide information on its shape and size, depending on the computed values. In contrast, the brightness distribution is characterized by skewness, which is quantified using third and higher-order moments.

Thus at the end of this process of phase retrieval and image reconstruction using the Training Set, we have three sets of images, which we may now refer to as the ``GT", ``N6" and ``N9" images respectively.

\section{Conclusion}

Intensity Interferometry (II) is re-emerging as a promising technique to overcome the challenges of very long baseline interferometry in the optical wavelength range.  However, compared to radio-interferometry, optical interferometry faces a major hurdle: photon correlation captures only the magnitude of the interferometric signal, resulting in a loss of phase information.

This work addresses the challenge of phase retrieval in II using a machine-learning technique, specifically a conditional Generative Adversarial Network (cGAN). Our study demonstrates that the size, shape and brightness distribution of fast rotating stars can be recovered by a Pix2Pix cGAN model trained on the combined input of the sky-image of known sources  along with their respective II data. In this training the sky-image acts as the real ``ground truth'' and the II data acts as the ``condition''. The Discriminator of our Model is trained to efficiently distinguish between the real images and fake (generated) images based on the ``ground truth'' images and the respective II data as the condition. The Generator is trained to generate progressively realistic images which are also consistent with the condition of the II data. The evaluation of the trained Model is then carried out by comparison of image moments of ground truth images and generated images. Specifically, the monopole, second, and third-order moments are compared. The results support the effectiveness of cGAN in achieving the stated objective.

While the results of this study highlight the significant potential of machine learning, and in particular the applicability of cGAN, for image reconstruction in II, several aspects require further refinement. First, an important factor in the reconstruction process is the extent of Fourier plane coverage, which depends on the number of available telescopes and the total observing time. The reasonable success of this piece of work suggests that greater coverage of the $(u,v)$ plane signal would play a critical role in projects of image reconstruction of more complicated stellar systems can be undertaken than higher number of baselines. Future work might explore different observatory layouts to assess their impact on image reconstruction quality.  Second, detector efficiencies, which impact the signal-to-noise ratio (SNR) of actual observational data, have not yet been incorporated; addressing these factors will be crucial for more accurate SNR estimation.  Third, exploring and comparing alternative methods for image generation could reveal approaches that outperform cGAN in reconstructing stellar images with II. Fourth, experimenting with different loss functions could provide additional insights into the reconstruction quality. Although further testing is needed to refine the GAN and enhance its robustness and reliability, our findings suggest that machine learning is a promising approach for phase reconstruction in II.

\section*{Note on software}

The code used for this work is available on the DOI:\hfil\break
\href{https://doi.org/10.5281/zenodo.17598807}{10.5281/zenodo.17598807}

\section*{Acknowledgements}
One of the authors (SS) gratefully acknowledges the computing facilities and the local hospitality extended to him by the Inter-University Centre for Astronomy and Astrophysics (IUCAA), Pune, India under its Visiting Associate Programme during the preparation and finalization of this manuscript.

\bibliographystyle{aa}
\bibliography{main}

@article{10.1093/mnras/stab2391,
	author = {Rai et al., Km Nitu},
	title = "{Radius measurement in binary stars: simulations of intensity interferometry}",
	journal = {Monthly Notices of the Royal Astronomical Society},
	volume = {507},
	number = {2},
	pages = {2813-2824},
	year = {2021},
	month = {08},
	issn = {0035-8711},
	doi = {10.1093/mnras/stab2391}
}

@article{Rai2025,
  author       = {Rai, K. Nitu and Basak, Soumen and Sarangi, Subrata and Saha, Prasenjit},
  title        = {Interference with (Pseudo) Thermal Light},
  journal      = {Resonance},
  year         = {2025},
  volume       = {30},
  pages        = {45--57},
  doi          = {10.1007/s12045-025-1729-x},
  publisher    = {Springer}
}

@article{10.1093/mnras/stac2433,
	author = {Rai et al., Km Nitu},
	title = "{Simulations of astrometric planet detection in Alpha Centauri by intensity interferometry}",
	journal = {Monthly Notices of the Royal Astronomical Society},
	volume = {516},
	number = {2},
	pages = {2864-2875},
	year = {2022},
	month = {08},
	issn = {0035-8711},
	doi = {10.1093/mnras/stac2433}
}

@ARTICLE{2013APh....43..331D,
       author = {{Dravins et al.}, Dainis},
        title = "{Optical intensity interferometry with the Cherenkov Telescope Array}",
      journal = {Astroparticle Physics},
         year = 2013,
        month = mar,
       volume = {43},
        pages = {331-347},
          doi = {10.1016/j.astropartphys.2012.04.017},
archivePrefix = {arXiv},
       eprint = {1204.3624},
 primaryClass = {astro-ph.IM}
}

@article{Haubois2009,
  author = {Haubois et al., X.},
  title = {Imaging the spotty surface of Betelgeuse in the H band},
  journal = {Astronomy \& Astrophysics},
  volume = {508},
  number = {2},
  pages = {923--932},
  year = {2009},
  doi = {10.1051/0004-6361/200912927},
  url = {https://doi.org/10.1051/0004-6361/200912927}
}

@article{Norris2021AZCyg,
  author       = {Norris et al., Ryan P.},
  title        = {Long Term Evolution of Surface Features on the Red Supergiant {AZ Cyg}},
  journal      = {The Astrophysical Journal},
  year         = {2021},
  volume       = {919},
  pages        = {124},
  doi          = {10.3847/1538-4357/ac0c7e},
}

@article{Liu2024SuperresolutionII,
  title        = {Super-resolution imaging based on active optical intensity interferometry},
  author       = {Liu, Lu-Chuan and Wu, Cheng and Li, Wei and Chen, Yu-Ao and Wilczek, Frank and Shao, Xiao-Peng and Xu, Feihu and Zhang, Qiang and Pan, Jian-Wei},
  journal      = {arXiv preprint},
  year         = {2024},
  volume       = {arXiv:2404.15685},
  doi          = {10.48550/arXiv.2404.15685},
  url          = {https://arxiv.org/abs/2404.15685},
}

@article{Liu2025,
  author = {Liu et al., Lu-Chuan},
  title = {Active Optical Intensity Interferometry},
  journal = {Physical Review Letters},
  volume = {134},
  number = {18},
  pages = {180201},
  year = {2025},
  doi = {10.1103/PhysRevLett.134.180201},
  url = {https://doi.org/10.1103/PhysRevLett.134.180201}
}

@article{goodfellow2014generative,
	title={Generative adversarial nets},
	author={Goodfellow et al., Ian},
	journal={Advances in neural information processing systems},
	volume={27},
	year={2014}
}

@article{goldberger1963use,
	title={Use of intensity correlations to determine the phase of a scattering amplitude},
	author={Goldberger et al., Marvin L},
	journal={Physical Review},
	volume={132},
	number={6},
	pages={2764},
	year={1963},
	publisher={APS}	
}

@article{gamo1963triple,
	title={Triple correlator of photoelectric fluctuations as a spectroscopic tool},
	author={Gamo, Hideya},
	journal={Journal of Applied Physics},
	volume={34},
	number={4},
	pages={875--876},
	year={1963},
	publisher={American Institute of Physics}
}

@article{GerchbergSaxton1972,
	title={A practical algorithm for the determination of phase from image and diffraction plane pictures},
	author={R. W. Gerchberg},
	journal={Optik},
	year={1972},
	volume={35},
	pages={237-246}
}

@article{Fienup1982,
	title={Phase retrieval algorithms: a comparison},
	author={Fienup, J.R.},
	journal={Applied Optics},
	volume={21},
	number={15},
	pages={2758-2769},
	year={1982}
}

@article{sato1978imaging,
	title={Imaging system using an intensity triple correlator},
	author={Sato et al., Takuso},
	journal={Applied optics},
	volume={17},
	number={13},
	pages={2047--2052},
	year={1978},
	publisher={Optica Publishing Group}
}

@article{sato1979computer,
	title={Computer controlled image sensor and its application},
	author={Sato et al., Takuso},
	journal={Applied Optics},
	volume={18},
	number={4},
	pages={485--488},
	year={1979},
	publisher={Optica Publishing Group}
}

@article{sato1981adaptive,
	title={Adaptive techniques for precise detection of the coherence function by an intensity triple correlator},
	author={Sato et al., Takuso},
	journal={Applied Optics},
	volume={20},
	number={12},
	pages={2055--2059},
	year={1981},
	publisher={Optica Publishing Group}
}

@inproceedings{holmes2010two,
	title={Two-dimensional image recovery in intensity interferometry using the Cauchy-Riemann relations},
	author={Holmes et al., RB},
	booktitle={Adaptive Coded Aperture Imaging, Non-Imaging, and Unconventional Imaging Sensor Systems II},
	volume={7818},
	pages={175--185},
	year={2010},
	organization={SPIE}
}

@article{LeBohec2006,
	title={Optical Intensity Interferometry with Atmospheric Cerenkov Telescope Arrays},
	author={Le Bohec, S; Holder, J.},
	journal={The Astrophysical Journal},
	volume={649},
	pages={399--405},
	year={2006},
	publisher={The American Astronomical Society.}
}

@inproceedings{nunez2010stellar,
	title={Stellar intensity interferometry: Imaging capabilities of air Cherenkov telescope arrays},
	author={Nu{\~n}ez et al., Paul D},
	booktitle={Optical and Infrared Interferometry II},
	volume={7734},
	pages={458--467},
	year={2010},
	organization={SPIE}
}

@inproceedings{Li2014,
	title={Phase retrieval using regularization method in intensity correlation imaging},
	author={Li et al., Xiyu},
	booktitle={Imaging Spectroscopy, Telescopes and Large Optics},
	volume={9298},
	pages={92981G1-G7},
	year={2014},
	organization={SPIE}
}

@article{nunez2012high,
	title={High angular resolution imaging with stellar intensity interferometry using air Cherenkov telescope arrays},
	author={Nu{\~n}ez et al., Paul D},
	journal={Monthly Notices of the Royal Astronomical Society},
	volume={419},
	number={1},
	pages={172--183},
	year={2012},
	publisher={The Royal Astronomical Society}
}

@article{nunez2012imaging,
	title={Imaging submilliarcsecond stellar features with intensity interferometry using air Cherenkov telescope arrays},
	author={Nu{\~n}ez et al., Paul D},
	journal={Monthly Notices of the Royal Astronomical Society},
	volume={424},
	number={2},
	pages={1006--1011},
	year={2012},
	publisher={Blackwell Science Ltd Oxford, UK}
}

@article{Teague1983,
	title={Deterministic phase retrieval: a Green's function solution},
	author={Teague, Michael R.},
	journal={Journal of Opticala Society of America},
	volume={73},
	number={11},
	pages={1434--1441},
	year={1983}
}

@article{Zhang2020,
	author = {Zhang et al., Jialin},
	journal = {Opt. Lett.},
	number = {13},
	pages = {3649--3652},
	publisher = {Optica Publishing Group},
	title = {On a universal solution to the transport-of-intensity equation},
	volume = {45},
	month = {Jul},
	year = {2020}
}

@article{Kirisits2024,
	title={Using the Transport of Intensity and the Transport of Phase Equation for Phase Retrieval},
	author={Kirisits et al., Clemens},
	journal={arXiv preprint arXiv:2406.14143v2},
	year={2024}
}

@article{hanbury1974angular,
	title={The angular diameters of 32 stars},
	author={Hanbury Brown et al., R},
	journal={Monthly Notices of the Royal Astronomical Society},
	volume={167},
	number={1},
	pages={121--136},
	year={1974},
	publisher={Oxford University Press Oxford, UK}
}

@article{HBT56Lab, 
   title={Correlation between photons in two coherent beams of light},
   author={Hanbury Brown, R and Twiss, Richard Q},
   journal={Nature},
   volume={177},
   number={4497},
   pages={27-29},
   year={1956},
   publisher={Nature Publishing Group UK London}
}

@article{brown1957interferometry,
  title={Interferometry of the intensity fluctuations in light-I. Basic theory: the correlation between photons in coherent beams of radiation},
  author={Hanbury Brown et al., R},
  journal={Proceedings of the Royal Society of London. Series A. Mathematical and Physical Sciences},
  volume={242},
  number={1230},
  pages={300--324},
  year={1957},
  publisher={The Royal Society London}
}

@article{brown1958interferometry,
  title={Interferometry of the intensity fluctuations in light. II. An experimental test of the theory for partially coherent light},
  author={Hanbury Brown et al., R},
  journal={Proceedings of the Royal Society of London. Series A. Mathematical and Physical Sciences},
  volume={243},
  number={1234},
  pages={291--319},
  year={1958},
  publisher={The Royal Society London}
}

@article{glauber1963quantum,
  title={The quantum theory of optical coherence},
  author={Glauber, Roy J},
  journal={Physical Review},
  volume={130},
  number={6},
  pages={2529},
  year={1963},
  publisher={APS}
}

@article{acciari2020optical,
	title={Optical intensity interferometry observations using the MAGIC imaging atmospheric Cherenkov telescopes},
	author={Acciari et al., V A},
	journal={MNRAS},
	volume={491},
	number={2},
	pages={1540--1547},
	year={2020},
	publisher={Oxford University Press}
}

@article{aleksic2016major,
	title={The major upgrade of the MAGIC telescopes, Part II: A performance study using observations of the Crab Nebula},
	author={Aleksi{\'c} et al., Jelena},
	journal={Astroparticle Physics},
	volume={72},
	pages={76--94},
	year={2016},
	publisher={Elsevier}
}

@article{mirza2014conditional,
	title={Conditional generative adversarial nets},
	author={Mirza et al., Mehdi},
	journal={arXiv preprint arXiv:1411.1784},
	year={2014}
}

@inproceedings{isola2017image,
	title={Image-to-image translation with conditional adversarial networks},
	author={Isola et al., Phillip},
	booktitle={Proceedings of the IEEE conference on computer vision and pattern recognition},
	pages={1125--1134},
	year={2017}
}

@inproceedings{ronneberger2015u,
	title={U-net: Convolutional networks for biomedical image segmentation},
	author={Ronneberger et al., Olaf},
	booktitle={Medical image computing and computer-assisted intervention--MICCAI 2015: 18th international conference, Munich, Germany, October 5-9, 2015, proceedings, part III 18},
	pages={234--241},
	year={2015},
	organization={Springer}
}

@article{schawinski2017galaxypics,
title={Generative adversarial networks recover features in astrophysical images of galaxies beyond the deconvolution limit},
author={Schawinski et al., Kevin},
	journal={MNRAS},
	volume={467},
	pages={L110--L114},
	year={2017},
	publisher={Oxford University Press}
}

@article{mustafa2019cosmogan,
title={CosmoGAN: creating high-fidelity weak lensing convergnece maps using Generative Adversarial Networks},
author={Mustafa et al., Mustafa},
	journal={Computational Astrophysics and Cosmology},
	volume={6},
	article=[1],
	pages={1--13},
	year={2019},
	publisher={Oxford University Press}
}

@article{coccomini2021lightweightgan,
title={Generatiive Adversarial Networks for Astronomical Images Generation},
author={Coccomini et al., Davide},
journal={arXiv preprint arXiv:2122.11578},
year={2021}
}

@article{abadi2016tensorflow,
	title={Tensorflow: Large-scale machine learning on heterogeneous distributed systems},
	author={Abadi et al., Mart{\'\i}n},
	journal={arXiv preprint arXiv:1603.04467},
	year={2016}
}

@article{virtanen2020scipy,
	title={SciPy 1.0: fundamental algorithms for scientific computing in Python},
	author={Virtanen et al., Pauli},
	journal={Nature methods},
	volume={17},
	number={3},
	pages={261--272},
	year={2020},
	publisher={Nature Publishing Group}
}

@article{4160265,
	title={Matplotlib: A 2D Graphics Environment}, 
	author={Hunter, John D.},
	journal={Computing in Science \& Engineering}, 
	year={2007},
	volume={9},
	number={3},
	pages={90-95},
	doi={10.1109/MCSE.2007.55}
}

@book{murphy2022probabilistic,
	title={Probabilistic machine learning: an introduction},
	author={Murphy, Kevin P},
	year={2022},
	publisher={MIT press}
}

@book{prince2023understanding,
	title={Understanding deep learning},
	author={Prince, Simon JD},
	year={2023},
	publisher={MIT press}
}

@article{hu1962visual,
	title={Visual pattern recognition by moment invariants},
	author={Hu, Ming-Kuei},
	journal={IRE transactions on information theory},
	volume={8},
	number={2},
	pages={179--187},
	year={1962},
	publisher={IEEE}
}

@article{teague1980image,
	title={Image analysis via the general theory of moments},
	author={Teague, Michael Reed},
	journal={Josa},
	volume={70},
	number={8},
	pages={920--930},
	year={1980},
	publisher={Optica Publishing Group}
}

@article{von1924radiative,
	title={The radiative equilibrium of a rotating system of gaseous masses},
	author={Von Zeipel, H},
	journal={Monthly Notices of the Royal Astronomical Society, Vol. 84, p. 665-683},
	volume={84},
	pages={665--683},
	year={1924}
}

@ARTICLE{1999A&A...347..185M,
	title = {Stellar evolution with rotation IV: von Zeipel's theorem and anisotropic losses of mass and angular momentum},
	author = {{Maeder}, Andr{\'e}},
	journal = {Astronomy \& Astrophysics},
	year = 1999,
	month = jul,
	volume = {347},
	pages = {185-193}
}

@article{lucy1967gravity,
	title={Gravity-darkening for stars with convective envelopes},
	author={Lucy, Leon B},
	journal={Zeitschrift f{\"u}r Astrophysik, Vol. 65, p. 89},
	volume={65},
	pages={89},
	year={1967}
}

@article{mcalister2005first,
	title={First results from the CHARA array. I. An interferometric and spectroscopic study of the fast rotator $\alpha$ Leonis (Regulus)},
	author={McAlister et al., Harold A},
	journal={The Astrophysical Journal},
	volume={628},
	number={1},
	pages={439},
	year={2005},
	publisher={IOP Publishing}
}

@ARTICLE{2020MNRAS.498.4577B,
	   title = "{Towards a polarization prediction for LISA via intensity interferometry}",
       author = {{Baumgartner et al.}, Sandra},
      journal = {MNRAS},
         year = 2020,
        month = nov,
       volume = {498},
       number = {3},
        pages = {4577-4589},
        primaryClass = {astro-ph.IM}
}

@ARTICLE{2024ApJ...966...28A,
	   title = "{An Angular Diameter Measurement of {\ensuremath{\beta}} UMa via Stellar Intensity Interferometry with the VERITAS Observatory}",
       author = {{Acharyya et al.}, A.},
      journal = {The Astrophysical Journal},
         year = 2024,
        month = may,
       volume = {966},
       number = {1},
          eid = {28},
        pages = {28},
        primaryClass = {astro-ph.SR}
}

@ARTICLE{2024MNRAS.529.4387A,
       author = {{Abe et al.}, S.},
        title = "{Performance and first measurements of the MAGIC stellar intensity interferometer}",
      journal = {MNRAS},
         year = 2024,
        month = apr,
       volume = {529},
       number = {4},
        pages = {4387-4404},
 primaryClass = {astro-ph.IM}
}

@ARTICLE{2025MNRAS.537.2334V,
       author = {{Vogel et al.}, Naomi},
        title = "{Simultaneous two-colour intensity interferometry with H.E.S.S}",
      journal = {MNRAS},
         year = 2025,
        month = mar,
       volume = {537},
       number = {3},
        pages = {2334-2341},
 primaryClass = {astro-ph.IM}
}

@ARTICLE{Archer-arXiv-2025,
       author = {{Archer}, A. and {Aufdenberg}, J.~P. and {Bangale}, P. and {Bartkoske}, J.~T. and {Benbow}, W. and {Buckley}, J.~H. and {Chen}, Y. and {Chin}, N.~B.~Y. and {Christiansen}, J.~L. and {Chromey}, A.~J. and {Duerr}, A. and {Escobar Godoy}, M. and {Feldman}, S. and {Feng}, Q. and {Filbert}, S. and {Fortson}, L. and {Furniss}, A. and {Hanlon}, W. and {Hervet}, O. and {Hinrichs}, C.~E. and {Holder}, J. and {Hughes}, Z. and {Humensky}, T.~B. and {Jin}, W. and {Johnson}, M.~N. and {Kertzman}, M. and {Kherlakian}, M. and {Kieda}, D. and {Korzoun}, N. and {LeBohec}, T. and {Lisa}, M.~A. and {Lundy}, M. and {Maier}, G. and {Matthews}, N. and {Moriarty}, P. and {Mukherjee}, R. and {Ning}, W. and {Ong}, R.~A. and {Pandey}, A. and {Pohl}, M. and {Pueschel}, E. and {Quinn}, J. and {Rabinowitz}, P.~L. and {Ragan}, K. and {Reynolds}, P.~T. and {Ribeiro}, D. and {Roache}, E. and {Rose}, J.~G. and {Sadeh}, I. and {Saha}, L. and {Santander}, M. and {Scott}, J. and {Sembroski}, G.~H. and {Shang}, R. and {Tak}, D. and {Tucci}, J.~V. and {Valverde}, J. and {Vassiliev}, V.~V. and {Williams}, D.~A. and {Wong}, S.~L. and {The VERITAS Collaboration}},
        title = "{Measurement of the photosphere oblateness of $γ$ Cassiopeiae via Stellar Intensity Interferometry with the VERITAS Observatory}",
      journal = {arXiv e-prints},
         year = 2025,
        month = jun,
          eid = {arXiv:2506.15027},
        pages = {arXiv:2506.15027},
          doi = {10.48550/arXiv.2506.15027},
archivePrefix = {arXiv},
       eprint = {2506.15027},
 primaryClass = {astro-ph.SR},
       adsurl = {https://ui.adsabs.harvard.edu/abs/2025arXiv250615027A}
}

@article{vanBelle2001,
  author = {van Belle, Gerard T. and Ciardi, David R. and Thompson, Robert R. and Akeson, Rachel L. and Lada, Elizabeth A.},
  title = {Altair's Oblateness and Rotation Velocity from Long-Baseline Interferometry},
  journal = {The Astrophysical Journal},
  volume = {559},
  number = {2},
  pages = {1155--1164},
  year = {2001},
  month = oct,
  doi = {10.1086/322340}
}

@article{DomicianodeSouza2003,
  author = {Domiciano de Souza, A. and Kervella, P. and Jankov, S. and Abe, L. and Vakili, F. and di Folco, E. and Paresce, F.},
  title = {The spinning-top Be star Achernar from VLTI-VINCI},
  journal = {Astronomy and Astrophysics},
  volume = {407},
  pages = {L47--L50},
  year = {2003},
  month = aug,
  doi = {10.1051/0004-6361:20030868}
}

@article{DomicianodeSouza2005,
  author = {Domiciano de Souza, A. and Kervella, P. and Jankov, S. and Vakili, F. and Ohishi, N. and Nordgren, T. E. and Abe, L.},
  title = {Gravitational-darkening of Altair from interferometry},
  journal = {Astronomy and Astrophysics},
  volume = {442},
  number = {2},
  pages = {567--578},
  year = {2005},
  month = nov,
  doi = {10.1051/0004-6361:20042476}
}

@article{Monnier2007,
  author = {Monnier, J. D. and Zhao, M. and Pedretti, E. and Thureau, N. and Ireland, M. and Muirhead, P. and Berger, J.-P. and Millan-Gabet, R. and Van Belle, G. and ten Brummelaar, T. and McAlister, H. and Ridgway, S. T. and Turner, N. and Sturmann, L. and Sturmann, J. and Berger, D. H.},
  title = {Imaging the Surface of Altair},
  journal = {Science},
  volume = {317},
  number = {5836},
  pages = {342--345},
  year = {2007},
  month = jul,
  doi = {10.1126/science.1143205}
}

@article{Pedretti2009,
  author = {Pedretti, E. and Monnier, J. D. and ten Brummelaar, T. and Thureau, N. D.},
  title = {Imaging with the CHARA interferometer},
  journal = {New Astronomy Reviews},
  volume = {53},
  number = {11-12},
  pages = {353--362},
  year = {2009},
  month = nov,
  doi = {10.1016/j.newar.2010.07.005}
}

@article{Zhao2009,
  author = {Zhao, M. and Monnier, J. D. and Pedretti, E. and Thureau, N. and M{\'e}rand, A. and ten Brummelaar, T. and McAlister, H. and Ridgway, S. T. and Turner, N. and Sturmann, J. and Sturmann, L. and Goldfinger, P. J. and Farrington, C.},
  title = {Imaging and Modeling Rapidly Rotating Stars: {$\alpha$} Cephei and {$\alpha$} Ophiuchi},
  journal = {The Astrophysical Journal},
  volume = {701},
  number = {1},
  pages = {209--224},
  year = {2009},
  month = aug,
  doi = {10.1088/0004-637X/701/1/209}
}

@article{Martinez2021,
  author = {Martinez, A. O. and Baron, F. and Monnier, J. D. and Roettenbacher, R. M. and Parks, J. R.},
  title = {ROTIR: A Tool for Parametric Modeling and Image Reconstruction of Stellar Surfaces from Interferometric Data},
  journal = {The Astrophysical Journal},
  volume = {916},
  number = {1},
  pages = {60},
  year = {2021},
  month = jul,
  doi = {10.3847/1538-4357/ac06a5}
}

@article{scuderi2022astri,
	title={The astri mini-array of cherenkov telescopes at the observatorio del teide},
	author={Scuderi, Salvatore and Giuliani, Andrea and Pareschi, Giovanni and Tosti, G and Catalano, O and Amato, E and Antonelli, LA and Gonzales, J Becerra and Bellassai, G and Bigongiari, C and others},
	journal={Journal of High Energy Astrophysics},
	volume={35},
	pages={52--68},
	year={2022},
	publisher={Elsevier},
	url = {https://www.sciencedirect.com/science/article/pii/S2214404822000180}
}

\end{document}